\documentclass[aps,prx,twocolumn,showpacs,floatfix, superscriptaddress]{revtex4-2}

\usepackage{subfigure}
\usepackage{psfrag,graphicx}
\usepackage{pict2e}
\usepackage{dcolumn}
\usepackage{amsmath,amssymb,braket}
\usepackage{bm}
\usepackage{color}
\usepackage{latexsym}
\usepackage{epstopdf}
\usepackage{color}
\usepackage{comment}
\usepackage[english]{babel}
\UseRawInputEncoding
\usepackage{amsfonts}
\usepackage{bm}
\usepackage{natbib}
\usepackage{bbold}

\usepackage{enumerate}

\usepackage{tikz}

\definecolor{myblue}{rgb}{0.2,0.2,0.8}
\definecolor{myzard}{cmyk}{0,0,0.05,0}
\definecolor{mywhite}{rgb}{1,1,1}
\definecolor{mywhite}{rgb}{1,1,1}
\definecolor{myred}{rgb}{1,0.,0.3}
\usepackage[colorlinks=true,citecolor=myblue,linkcolor=myred]{hyperref}

\definecolor{darkgreen}{rgb}{0.0, 0.4, 0.26}

\definecolor{mygrey}{gray}{0.35}
\definecolor{myblue}{rgb}{0.2,0.2,0.8}
\definecolor{myzard}{cmyk}{0,0,0.05,0}
\definecolor{mywhite}{rgb}{1,1,1}
\definecolor{mywhite}{rgb}{1,1,1}
\definecolor{myred}{rgb}{1,0.,0.3}

\def\be{\begin{equation}}
\def\ee{\end{equation}}
\def\ba{\begin{align}}
\def\enda{\end{align}}
\def\bi{\begin{itemize}}
\def\ei{\end{itemize}}

\def\beq{\begin{equation}}
\def\beq{\begin{equation}}
\def\eeq{\end{equation}}

\def\ii{{\bm i}}
\def\jj{{\bm j}}

\def\rr{{\bm r}}

\def\ss{{\textbf s}}

\newcommand{\mean}[1]{\langle #1\rangle}

\begin{document}

\title{Programming optical-lattice Fermi-Hubbard quantum simulators}

 \author{Cristian Tabares}
 \email{cristian.tabares@csic.es}
 \affiliation{Institute of Fundamental Physics IFF-CSIC, Calle Serrano 113b, 28006 Madrid, Spain}
\author{Christian Kokail}
  \affiliation{ITAMP, Harvard-Smithsonian Center for Astrophysics, Cambridge, MA 02138, USA}
 \affiliation{Department of Physics, Harvard University, Cambridge, MA 02138, USA}
\author{Peter Zoller}
 \affiliation{Institute for Theoretical Physics, University of Innsbruck, 6020 Innsbruck, Austria}
 \affiliation{Institute for Quantum Optics and Quantum Information of the Austrian Academy of Sciences, 6020 Innsbruck, Austria}
\author{Daniel~Gonz\'alez-Cuadra}
\email{dgonzalezcuadra@fas.harvard.edu}
 \affiliation{Institute of Fundamental Physics IFF-CSIC, Calle Serrano 113b, 28006 Madrid, Spain}
 \affiliation{Department of Physics, Harvard University, Cambridge, MA 02138, USA}
 \affiliation{Institute for Theoretical Physics, University of Innsbruck, 6020 Innsbruck, Austria}
 \affiliation{Institute for Quantum Optics and Quantum Information of the Austrian Academy of Sciences, 6020 Innsbruck, Austria}
 
 \author{Alejandro Gonz\'alez-Tudela}
 \email{a.gonzalez.tudela@csic.es}
 \affiliation{Institute of Fundamental Physics IFF-CSIC, Calle Serrano 113b, 28006 Madrid, Spain}
\begin{abstract}

Fermionic atoms in optical lattices provide a native implementation of Fermi-Hubbard (FH) models that can be used as analog quantum simulators of many-body fermionic systems. Recent experimental advances include the time-dependent local control of chemical potentials and tunnelings, and thus enable to operate this platform digitally as a programmable quantum simulator. Here, we explore these opportunities and develop ground-state preparation algorithms for different fermionic models, based on the ability to implement both single-particle and many-body, high-fidelity fermionic gates, as provided by the native FH Hamiltonian. In particular, we first design variational, pre-compiled quantum circuits to prepare the ground state of the natively implemented FH model, with significant speedups relative to competing adiabatic protocols. Besides, the versatility of this variational approach enables to target extended FH models, i.e., including terms that are not natively realized on the platform. As an illustration, we include next-nearest-neighbor tunnelings at finite dopings, relevant in the context of $d$-wave superconductivity. Furthermore, we discuss how to approximate the imaginary-time evolution using variational fermionic circuits, both as an alternative state-preparation strategy, and as a subroutine for the Quantum Lanczos algorithm to further improve the energy estimation. We benchmark our protocols for ladder geometries, though they can be readily applied to 2D experimental setups to address regimes beyond the capabilities of current classical methods. These results pave the way for more efficient and comprehensive explorations of relevant many-body phases with existing programmable fermionic quantum simulators.
\end{abstract}

\maketitle


\section{Introduction}~\label{sec:system}

Ultracold atoms in optical lattices are now established as one of the leading platforms to simulate quantum many-body physics~\cite{Jaksch_2005, Lewenstein_2007, Gross_2017, Schafer_2020}. The use of fermionic atoms, in particular, has enabled the study of strongly-correlated fermionic problems in condensed-matter~\cite{Hart_2015, Boll_2016, Cheuk_2016, Mazurenko_2017, Xu_2023, Shao_2024,Chiu_2019, Bourgund_2023,Sompet_2022,Hirthe_2023, Hartke_2023,Schreiber_2015, Nichols_2019, Guardado-Sanchez_2020, Scherg_2021} and high-energy physics~\cite{Banuls_2020, Aidelsburger2022, DiMeglio_2024}. Traditionally, cold-atom quantum simulators operate in an analogue mode, where the system evolves globally under the local Fermi-Hubbard (FH) Hamiltonian, consisting on nearest-neighbor (NN) tunneling terms and on-site Hubbard interactions. This analogue approach complements current efforts on digital simulation with qubit-based devices~\cite{Google2020_HF,arute2020observationseparateddynamicscharge,Stanisic2022,Tazhigulov2022,Hemery2024,Clinton2024,robledomoreno2024chemistryexactsolutionsquantumcentric,Evered_2025}, with the advantage of avoiding the overheads of fermion-to-qubit mappings~\cite{Abrams_1997, Ortiz_2001, Bravyi_2002, Ball_2005, Verstraete_2005, Whitfield_2011, Whitfield_2016}, as well as showing a good performance against errors in the absence of quantum error correction~\cite{Preskill_2018, Daley_2022,Flannigan_2022errors,Trivedi_2023,schiffer2024quantumadiabaticalgorithmsuppresses,kashyap2024accuracyguaranteesquantumadvantage}. In certain regimes, however, ground-state preparation with fermionic cold-atom simulators presents open challenges. This is the case when the energy gap is small, where state preparation is limited by temperature effects and finite coherence times, as well as for models beyond the natively implemented local FH Hamiltonian. Addressing both of them is required to observe certain phenomena, such as d-wave superconductivity~\cite{Qin_2022}, which is now believed to appear in the presence of next-nearest-neighbor (NNN) tunneling terms~\cite{Qin2020,Xu_2024}.

Recent experimental advances, including the single-site control and measurement of tunneling elements and chemical potentials using digital mirror devices, optical tweezers and superlattices~\cite{Dai2016, Qiu_2020, Yang2020a,Yang2020b,Zhang2023_atoms,Chalopin_2025,Wei_2023, Impertro_2024}, are enabling a more \textit{programmable} mode of operation, allowing to implement FH dynamics acting only on certain parts of the system. In this context, fermionic atoms in optical lattices can be regarded as \textit{fermionic quantum processors}~\cite{Gonzalez-Cuadra_2023, Zache_2023, Gkritsis_2024, Schuckert_2024, Ott_2024}, i.e. digital quantum simulators with fermions as their basic constituents. The system's native dynamics are harnessed here as \textit{resources} to build a set of fermionic gates, which are then used to construct quantum-simulation circuits. This approach combines the error resilience of analog devices, since complicated many-body gates can be implemented directly with high fidelity, while incorporating the versatility of digital systems. The latter enables to prepare ground states more efficiently than traditional adiabatic protocols, and to \textit{target} more complicated Hamiltonians that are never directly implemented in the system.  While this programmable approach has been considered for trapped ions~\cite{Kokail2019, Monroe_2021, Joshi2023} and Rydberg atoms arrays~\cite{Labuhn_2016, Bernien_2017, Keesling_2019, Leseleuc_2019, Ebadi_2021, Scholl_2021, Bluvstein_2021, Semeghini_2021, Scholl_2022, Manovitz_2024, Lukin_2024, Gonzalez_2024}, the qubit-based algorithms developed for these platforms can not be applied directly to cold-atom simulators based on itinerant particles due to the fundamental differences in the hardware architecture and slower experimental repetition rates.

\begin{figure*}
    \centering
    \includegraphics[width=\linewidth]{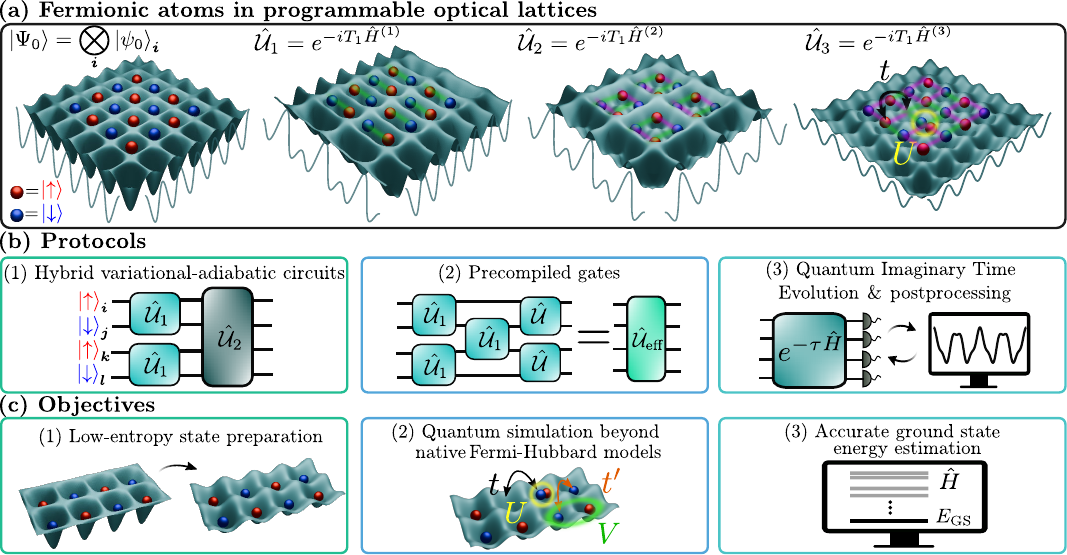
    }
    \caption{{\bf Schematic illustration of the programmable fermionic quantum simulator architecture introduced in this work.} (a) We consider a fermionic quantum processor based on ultracold (spinfull) fermionic atoms on a 2D optical superlattice. The superlattice structure is generated using two counter-propagating laser beams in each direction, giving rise to a four-site unit cell structure. Our protocols start by first preparing a Mott-insulating state, $\ket{\Psi_0}$, with atoms localized on the potential wells. In the left panel, we represent a half-filled state, but states with finite dopings can also be prepared. In this setup, different unitary gates $\hat{\mathcal{U}}_k\in\mathcal{G}$~\eqref{eq:Gateset} can be implemented by globally controlling the superlattice potential. Here, the atoms evolve under the FH Hamiltonian with tunnelings restricted to certain links, as schematically depicted in the other panels. (b) Different protocols built using these resources, and (c) applications in quantum simulation that we target in this work. (1) We construct hybrid variational-adiabatic circuits using sequences of programmable quenched evolutions, allowing us to prepare the ground state of different fermionic Hamiltonians starting from low-entropy product states.  (2) Our protocols can be also applied to extended FH models using only the resources depicted in (a), this is, from the native (local) FH dynamics. To achieve this, we either use the latter to construct precompiled gates that implement unitary evolutions under the extended terms or incorporate their effect through classical postprocessing. (3) The prepared states can be used as inputs to implement a Quantum Imaginary Time Evolution (QITE), and also to improve the estimation of the ground state energy using the Quantum Lanczos algorithm through classical post-processing.}
   \label{fig:1}
\end{figure*}

In this work, we focus on the problem of ground-state preparation for extended FH models relevant in condensed-matter physics, including NNN tunneling terms and NN interactions, and develop protocols especially tailored to the experimental capabilities of fermionic atoms in programmable optical lattices. For this, we first show how to implement a set of fermionic many-body gates generated by FH dynamics [Fig.~\ref{fig:1}(a)]. While not being universal, this gate set enables the construction of a broad class of state-preparation variational circuits~\cite{Cerezo_2021} [Fig.~\ref{fig:1}(b)]. In contrast to previous variational approaches applied to quantum chemistry problems using qubit-based devices~\cite{peruzzo14a,Kivlichan2018,Grimsley2019,Tang2021,Takeshita2020,Huggins2021}, the fermionic nature of our resources combined with the translational invariance of the target models allows us to construct circuits with few variational parameters, and to considerably reduce or even eliminate costly quantum-classical optimization loops.


The first class of fermionic circuits that we consider can be regarded as \textit{hybrid variational-adiabatic} circuits. Specifically, here the variational part is pre-complied classically to prepare the ground state on small system sizes, which are then adiabatically coupled to reach the ground state of the target extended system. This allows us to build finite correlation lengths much faster than purely adiabatic protocols, and at the same time avoid any quantum optimization on the hardware. 

The second class of variational circuits are inspired by the \textit{quantum imaginary-time evolution} (QITE)~\cite{motta2020a,McArdle2019,Yuan2019theoryofvariational,Gacon2024}, and can be interpreted as its unitary approximation. While in this case the optimization of the circuit is performed on the quantum hardware, we design our protocols to minimize experimental overheads. In particular, we investigate the effect of shot noise, and show that the number of experimental runs required to implement this algorithm is within reach for current cold-atom experiments. Finally, we show that this protocol has the additional advantage that the collected data can be post-processed using the \textit{Quantum Lanczos} method~\cite{motta2020a}, allowing us to further improve the estimation of the target ground-state energy.

The rest of the manuscript is structured as follows: we start in Sec.~\ref{sec:programmable} introducing a programmable quantum simulator based on fermionic atoms in optical lattices, including the fermionic gate set that can be implemented using the system's native dynamics. In Sec.~\ref{sec:state}, we build hybrid variational-adiabatic circuits for ground-state preparation using this gate set, and apply them to local and extended FH models, both at half filling as well as for finite dopings. In Sec.~\ref{sec:ITE}, we explain how to implement QITE using fermionic circuits, and apply it to approximate the ground state of the FH models, as well as to improve the estimation of the ground-state energy through the QLanczos algorithm. Finally, in Sec.~\ref{sec:conclusion} we summarize our findings and point to future directions.

\section{Programmable fermionic quantum simulator~\label{sec:programmable}}

\subsection{Resource Hamiltonians and fermionic gate set}

In this work, we consider a fermionic quantum processor based on fermionic atoms trapped in optical lattice potentials, with two internal levels that codify the spin degree of freedom, $\sigma \in \{\uparrow, \downarrow\}$ [Fig.~\ref{fig:1}(a)]. Assuming an optical lattice generated by counter-propagating laser beams, the dynamics of the atomic system can be described by the \textit{resource} local Fermi-Hubbard (FH) Hamiltonian,
\begin{align}~\label{eq:FH}
\hat{H}_{\mathrm{FH}}=\hat{H}_t+\hat{H}_U + \hat{H}_n\,\,,
\end{align}
consisting on nearest-neighbour tunneling terms,
\begin{align}
\hat{H}_t=-\sum_{\langle\ii,\jj\rangle,\sigma} t_{\ii,\jj}\hat{c}_{\ii,\sigma}^\dagger \hat{c}^{\vphantom{\dagger}}_{\jj,\sigma}\,,\label{eq:Ht}
\end{align}
on-site Hubbard interactions,
\begin{equation}
    \hat{H}_U = U \sum_{\ii}\hat{n}_{\ii,\uparrow}\hat{n}_{\ii,\downarrow}\,,\label{eq:HU}
\end{equation}
and local chemical potentials,
\begin{align}
    \hat{H}_n=\sum_{\ii,\sigma}\mu_{\ii,\sigma} \hat{n}_{\ii,\sigma}\,.
\end{align}
Here $c^{(\dagger)}_{\ii,\sigma}$ denote the (creation) annihilation fermionic operators corresponding to the $\sigma$ spin component at the site $\ii$ of a 2D square lattice, and $\hat{n}_{\ii,\sigma} = \hat{c}_{\ii,\sigma}^\dagger \hat{c}_{\ii,\sigma}$ is the atomic occupation number. 

To construct fermionic quantum-simulation circuits, we employ a set of unitary gates $\mathcal{G}$ generated by the native FH dynamics,
\begin{equation}
\mathcal{G} = \left\{e^{-i\left[\hat{H}_t(t_{\ii,\jj}) + H_U(U) + H_n(\mu_\ii)\right]T}\right\}_{t_{\ii,\jj},U,n_\ii} \,.\label{eq:Gateset} 
\end{equation}
In this work, we consider translational-invariant patterns for $t_{\ii,\jj}$ and $\mu_\ii$ [see Fig.~\ref{fig:1}(a)], implemented through superlattice potentials~\cite{Zhang2023_atoms,Impertro_2024, Chalopin_2025} and local energy shifts~\cite{Kaufman_2021, Young_2022, yan2022, Tao_2024}, respectively, where the latter can be made spin dependent by introducing a magnetic field. Together with a homogeneous Hubbard parameter $U$ tuned with Feschbach resonances, this gate set $\mathcal{G}$ can thus be implemented using only global experimental knobs. We will assume that we can turn on and off their corresponding parameters freely, and apply quench evolutions under the corresponding terms, which is what we label as programmable operation. 

We note that implementing these many-body fermionic gates using digital qubit devices would require long circuits, whose depths are limited in the absent of error correction, while here they are realized with high fidelity using the fermions' native dynamics. Despite not being universal, below we show how this restricted gate set is sufficient to construct different quantum state-preparation protocols for several translational-invariant fermionic Hamiltonians relevant in condensed-matter physics. 

\subsection{Initialization and state preparation}

For cold atoms in optical lattices acting as analogue quantum simulators, the ground state of the FH Hamiltonian~\eqref{eq:FH} can be prepared in the system using different approaches. The standard technique consists on first cooling down an atomic gas to ultracold temperatures, and then adiabatically ramping up the optical lattice, an approach that is limited by the temperatures that can be reached experimentally. Alternatively, one can take a more algorithmic approach by first preparing a low-entropy product state of fermionic atoms,
\begin{align}
\label{eq:product_state}
\ket{\Psi_0}=\bigotimes_{\ii} \ket{\psi_{0}}_\ii\,.
\end{align}
where $\ket{\psi_0}\in\{\ket{0}, \ket{\uparrow}, \ket{\downarrow}, \ket{\uparrow\downarrow\}}$ denotes the fermionic occupation on each lattice site. In particular, both the half-filled case, with $\rho = 1 / L \sum_{\ii,\sigma}\langle\hat{n}_{\ii,\sigma}\rangle = 1$, where $N$ is the number of lattice sites, as well as finite dopings, with $\delta = 1 - \rho\neq 0$, can be prepared using pattern-loading techniques~\cite{Rabl2003,Griessner_2007,Popp2006,Bakr2011}, as has been demonstrated experimentally~\cite{Dai2016,Yang2020a,Yang2020b,Zhang2023_atoms,Xu_2025}. These product states can be then adiabatically connected to the target state~\cite{Trebst2006, Xu_2025}. Although this approach is guaranteed to work if the ground state is gapped, it is limited by the coherence time of cold-atom experiments if the gap is small, that is the case of doped phases. Moreover, this approach can not be applied to target extended FH models whose interactions or tunnelings are not natively implemented in the system.

In the rest of the paper, we introduce different state-preparation protocols that address these open questions. We follow the algorithmic approach mentioned above, starting from low-entropy product states~\eqref{eq:product_state} that we take as a fermionic computational basis, but use instead variational circuits built from the fermionic gates in $\mathcal{G}$~\eqref{eq:Gateset} to prepare the target ground state [Fig.~\ref{fig:1}(b)]. This allows us to reach low entropies with a much shorter implementation time, as well as to target Hamiltonians with extended terms not included in the resource Hamiltonian [see, e.g., App.~\ref{app:extended_FH}]. As we explain in detail in the following sections, we design these variational algorithms to minimize the corresponding measurement overhead, tailoring them to the measurement capabilities of cold-atom quantum hardware, that we describe in App.~\ref{appendix:measurements}.

\section{Hybrid variational-adiabatic fermionic circuits~\label{sec:state}}

In this section, we design variational fermionic circuits with the resources described in Sec.~\ref{sec:programmable} to approximate the ground states of relevant condensed-matter Hamiltonians. To illustrate the power of these variational circuits, we consider first the local FH model $\hat{H}_{\mathrm{FH}}$ on a ladder geometry, and target a strongly-correlated regime with $U/t=8$, both at half filling [Sec.~\ref{subsec:half}] and finite dopings [Sec.~\ref{subsec:doping}]. In Sec.~\ref{subsec:extendedtun}, we show how these protocols can be also applied to extended FH models.

\subsection{Fermi-Hubbard ladders at half filling~\label{subsec:half}}
\begin{figure*}[tb]
    \centering
    \includegraphics[width=\linewidth]{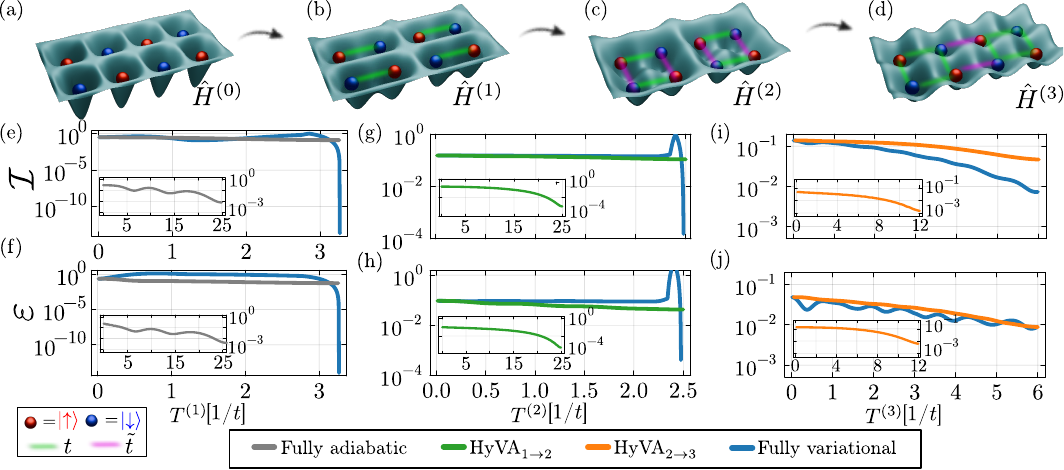}
    \caption{{\bf Ground-state preparation for the local FH model at half filling.} (a)--(d) Superlattice configurations employed to prepare the ground state of the local FH model on a $2 \times 4$ ladder at half filling and for $U/t = 8$, starting from an antiferromagnetic product state on a deep lattice (a). We employ quench evolutions under the corresponding Hamiltonians $\hat{H}^{(k)}$, where we depict the different non-zero tunneling elements in different colors. We first prepare the ground state on each uncoupled dimer (b), then for each uncoupled plaquettes (c), and finally for the full extended system (d). (e) and (f) show the infidelity $\mathcal{I}$ and the residual energy $\epsilon$, respectively, as a function of the implementation time $T^{(1)}$ to prepare a product of uncoupled dimers (b) from the initial product state in (a), using both variational and adiabatic protocols. The inset shows the results obtained with a longer adiabatic protocol. (g) and (h) show $\mathcal{I}$ and $\epsilon$ as a function of $T^{(2)}$ during the preparation of the ground state for a product of uncoupled plaquettes (c). Both protocols start from the state prepared variationally in (b), and $\mathrm{HyVA}_{1\rightarrow2}$ corresponds here to the adiabatic evolution of the latter to the target state. In the inset, we show results obtained for a longer adiabatic evolution. (i) and (j) show $\mathcal{I}$ and $\epsilon$ for the preparation of the ground state of the target Hamiltonian, i.e., the local FH model (d), as a function of $T^{(3)}$, starting from the state prepared in (c). In this case, the adiabatic part in the $\mathrm{HyVA}_{2\rightarrow3}$ protocol refers only to the adiabatic fusion of the uncoupled plaquettes, which are prepared with the fully variational protocol in (c). In the inset, we show again the results obtained after applying the HyVA protocol for longer times.} 
    \label{fig:3} 
\end{figure*}

The first target Hamiltonian is the local FH model~\eqref{eq:FH}, with $t_{\ii,\jj} = t$ and $\mu_\ii = 0$. This model has been regarded as a candidate to explain high$-T_c$ superconductivity in cuprate materials~\cite{Qin_2022}, although recent numerical evidence suggests that it does not support a dominant superconducting order in the expected parameter regime~\cite{Qin2020}. Nevertheless, the local FH model features other interesting strongly-correlated phases, such as stripe phases~\cite{Bourgund_2023}, away from half-filling.

As an initial state for our protocol, we consider an antiferromagnetic product state $\ket{\Psi_0}$ on a ladder geometry [see Fig.~\ref{fig:3}(a)]. This state is the half-filled ground state of the Hamiltonian
\begin{equation}
    \hat{H}^{(0)} = -\mu\sum_{\ii}\left(-1\right)^{i_x + i_y}\left(\hat{n}_{\ii,\uparrow}-\hat{n}_{\ii,\downarrow}\right)\,,
\end{equation}
with $\ii = (i_x, i_y)$. The steps of the protocol are schematically depicted in Figs.~\ref{fig:3}(b-d): i) we first prepare the ground state of $\hat{H}_\text{FH}$ for a product of uncoupled dimers; ii) we then couple neighboring dimers to obtain the ground state of $2\times 2$ plaquettes; and finally, iii) we obtain the ground states of $\hat{H}_\text{FH}$ for the whole system by coupling neighboring plaquettes. Each step $k$ of the protocol can be described as a variational quantum circuit consisting of a sequence of global unitary evolutions from our fermionic gate set $\mathcal{G}$~\eqref{eq:Gateset},
\begin{equation}
\label{eq:U_circuit}
\mathcal{\hat{U}}^{(k)}(\bm{\theta}^{(k)}) = \prod_{\mu = 1}^{D^{(k)}} e^{-\ii T^{(k)}_\mu \hat{H}^{(k)}(\bm{\lambda}_\mu^{(k)})}\, .
\end{equation}
Here $D^{(k)}$ denotes the depth of the circuit for each step $k$, $\bm{\theta}^{(k)} = (\bm{\theta}^{(k)}_1,\dots,\bm{\theta}^{(k)}_D)$ are variational parameters, and $\hat{H}^{(k)}$ are the programmable FH Hamiltonian depicted in Fig.~\ref{fig:3}, and written explicitly in App.~\ref{app:Hamiltonians}. For each step, we keep certain Hamiltonian couplings fixed and use other ones, denoted by the vector $\bm{\lambda}^{(k)}_\mu$, together with the evolution times $T^{(k)}_{\mu}$, as variational parameters $\bm{\theta}^{(k)}_\mu = T^{(k)}_{\mu}\bm{\lambda}^{(k)}_\mu$. The latter are optimized to minimize the expectation value of the target Hamiltonian over the prepared state using the variational quantum eigensolver (VQE) algorithm~\cite{Cerezo_2021}. 

In the following, we use as figures of merit to benchmark the protocols both the residual energy $\varepsilon$ and the infidelity $\mathcal{I}$ defined as
\begin{align}
    \label{eq:residualene}
    \varepsilon &= \frac{|E(\bm{\theta}^{(k)}) - E_\mathrm{GS}|}{N}\,,\\
    \label{eq:infidelity}
    \mathcal{I} &= 1 - \mathcal{F},
\end{align}
where $E(\bm{\theta}^{(k)}) = \braket{\psi(\bm{\theta}^{(k)})|\hat{H}_{\mathrm{FH}}|\psi(\bm{\theta}^{(k)}}$ is the expectation value of $\hat{H}_{\mathrm{FH}}$ over the variational state after each step $|\psi(\bm{\theta}^{(k)})\rangle=\mathcal{\hat{U}}^{(k)}\left(\bm{\theta}^{(k)}\right)|\psi(\bm{\theta}^{(k-1)})\rangle$, $E_{\mathrm{GS}}$ is the exact ground state energy, and $\mathcal{F} = |\langle \psi_\mathrm{GS}|\psi(\bm{\theta}^{(k)})\rangle\,|$ is the fidelity between the variational  and the exact ground state $|\psi_{\mathrm{GS}}\rangle$. We note that $E(\bm{\theta}^{(k)})$ can be measured experimentally using quantum gas microscopes~\cite{Bakr_2009, Sherson_2010, Gross_2021, Mark_2024}, as we explain in App.~\ref{appendix:measurements}. 

As we describe now in detail, in the first two steps a short circuit will be sufficient given the small system size of the target states, and we can thus optimize the circuit classically and apply the resulting optimal evolution directly on the quantum hardware. We refer to these variational circuits as variational \textit{pre-compiled}, to distinguish them from the ones required in step (iii), where the optimization should be performed on the quantum hardware, or replaced by an adiabatic fusion.

In Fig.~\ref{fig:3}(e)-(f), we show the results for the first step of the protocol, where we prepare a product of dimers. We compare the results obtained for an optimized variational circuit with $D^{(1)}=3$ with an adiabatic fusion of the dimers, and plot the evolution of $\varepsilon$ and $\mathcal{I}$ up to the total evolution time, $T^{(1)}=\sum_{\mu} T^{(1)}_{\mu}$. We note that the finite-depth variational circuit prepares \emph{exactly} the ground-state of the FH Hamiltonian on each uncoupled dimer. Compared to this, an adiabatic protocol interpolating from $\hat{H}^{(0)}$ to $\hat{H}_\text{FH}$ using the same implementation time does not reach similar accuracies, where one fifth of the total time is used to ramp up the values of $t$ and $U$ from zero, and the rest is used to decrease $\mu$. As shown in the inset, by increasing the adiabatic implementation time to $T = 25/t$, the resulting state is still limited to $\mathcal{I} \sim 10^{-2}$ and $\varepsilon \sim 10^{-3}$. We thus conclude that the pre-compiled variational circuit prepares the state with a much higher accuracy, requiring moreover a much shorter implementation time.

The results obtained for the second step of the protocol are shown in Fig.~\ref{fig:3}(g)-(h), where we prepare the ground state of a product of uncoupled plaquettes starting from the dimer product state, taking again $D^{(2)}=3$. We now compare these results with those obtained using a hybrid protocol, where we start from the same state, but then implement an adiabatic linear ramp of $\tilde{t}$ from $0$ to $t$. We refer to this as a hybrid-variational adiabatic ($\mathrm{HyVA}_{1\rightarrow2}$) approach. In this case, the fully variational protocol provides much better results, as it prepares the ground state of the uncoupled plaquettes with $\mathcal{I}\sim 10^{-4}$, and with an implementation time ten times shorter. 

In the final step, we switch on the tunneling elements between different plaquettes to prepare the ground state of the complete $2\times 4$ ladder [Fig.~\ref{fig:3}(d)]. In this case, we use the intra-plaquette tunnelings $\tilde{t}_\mu$ and the total evolution time for each quench $T_\mu$ as variational parameters. In Fig.~\ref{fig:3}(i)-(j), we show the results for a circuit with $D^{(3)}=3$ quench evolutions, where we obtain a fidelity above $99.9\%$. We compare these results with those obtained using a $\mathrm{HyVA}_{2\rightarrow3}$ protocol, starting now from the same variationally prepared plaquettes, which are then adiabatically fused using a linear ramp for the plaquette coupling. In this case, the fully variational protocols requires half of the implementation time compared to the HyVA protocol to reach the same fidelities.

These results show that, even if the $\mathrm{HyVA}_{2\rightarrow3}$ protocol is a bit slower, the difference is not very large compared to the fully variational one. Moreover, the former has the additional advantage that it only relies on pre-compiled variational circuits to prepare the ground state of the uncoupled plaquettes, which can be optimized classically as we show in the second step. Thus, it can be implemented directly on the quantum hardware for an extended 2D system, as opposed to a fully variational method. The latter requires a quantum-classical optimization loop~\cite{Cerezo_2021}, which is challenging to implement in cold-atom setups due to slow experimental repetition rates. We note that the good performance of the $\mathrm{HyVA}_{2\rightarrow3}$ protocol relies on the finite correlation length of the target ground state, and it is reminiscent of a renormalization-group approach. The pre-compiled variational circuits efficiently build up such correlations on plaquettes using a much shorter evolution time than their adiabatic counterparts, such that the final ground state can be then efficiently prepared by adiabatically coupling them. In the next section, we investigate the performance of this protocol as a function of the system size, showing how this advantage is maintained for larger systems. In the following, we drop the subindices and refer to this last protocol as HyVA.

\begin{figure}[tb]
    \centering
        \includegraphics[width=\linewidth]{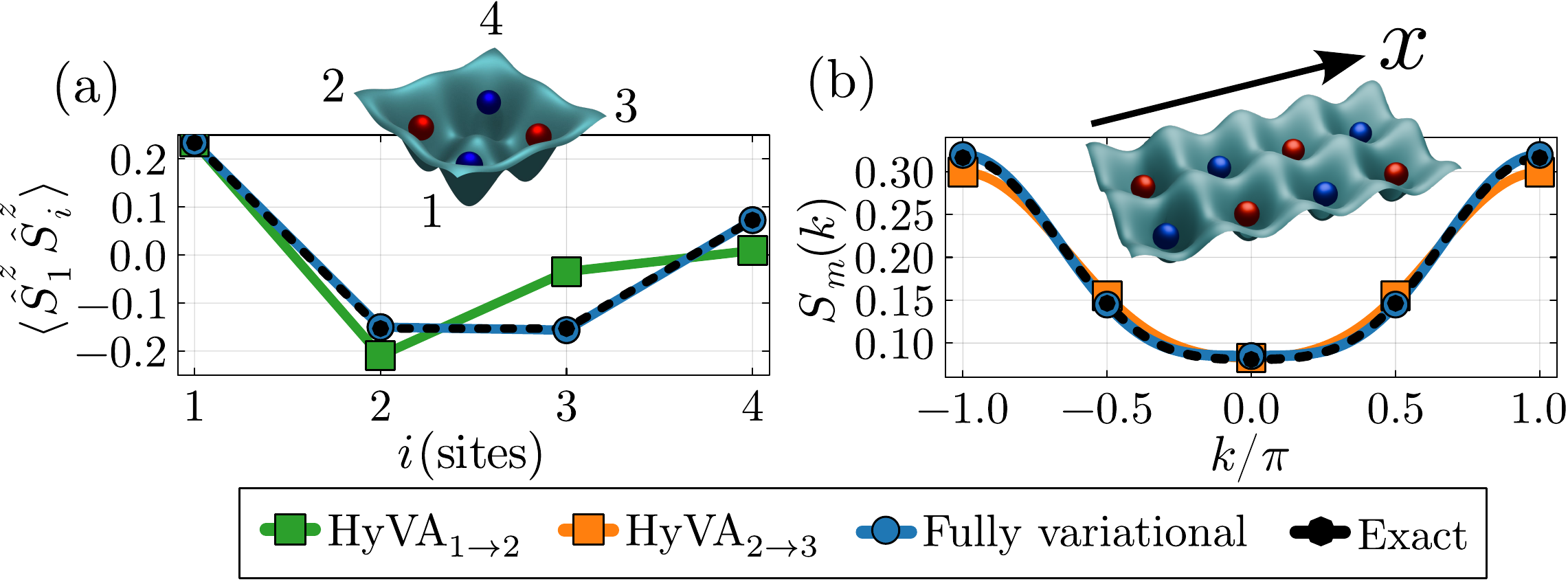}
    \caption{{\bf Observables for the prepared states.} (a) Spin-spin correlations between the atoms forming one (uncoupled) plaquette, measured on the state prepared after the second step of our protocol [Fig.~\ref{fig:3}(c)]. We compare the results with the fully variational circuit (in blue) and the $\mathrm{HyVA}_{1\rightarrow2}$ approach (in orange) with the same total implementation time. (b) Spin structure factor $S_m(k)$ along the $x$ axis for a ground state prepared on the $2\times 4$ ladder [Fig.~\ref{fig:3}(c)], obtained at the end of the third step using either a fully variational or the corresponding $\mathrm{HyVA}_{2\rightarrow3}$ protocol (green), using again the same implementation time. In both panels, we include the result for the exact ground state (in black).}
    \label{fig:4}
\end{figure}

\begin{figure*}[tb]
    \centering
    \includegraphics[width=0.95\linewidth]{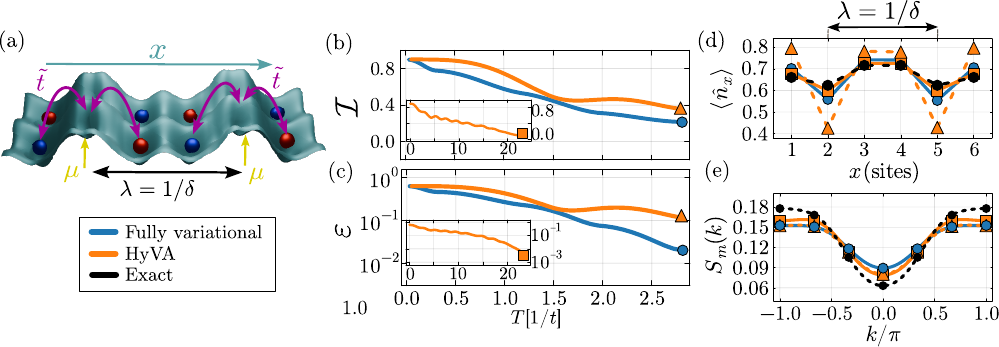}
    \caption{{\bf Ground-state preparation for the local FH model at finite doping.} (a) Initial state used to prepared the ground state of a $2\times6$ FH ladder with doping $\delta = 1/3$ and $U/t = 8$. Atoms are loaded into the optical lattice, with certain sites left empty according to the expected stripe configuration in the ground state, which has a wavelength $\lambda = 1/\delta$. (b) and (c) show the infidelity $\mathcal{I}$ and residual energy $\varepsilon$, respectively, as a function of the implementation time $T$. We compare the results obtained with the variational circuit (blue) in~\eqref{eq:U_links_delta} with the HyVA protocol (orange) described in the main text. The insets show the results obtained with the latter for a longer adiabatic evolution time. (d) and (e) show the charge density profile $\langle\hat{n}_x\rangle$ and spin structure factor along the $x-$axis $S_m(k)$, respectively, for the final states prepared with these different methods. Markers indicate the observables are obtained for the different times indicated in panels (b) and (c).} 
    \label{fig:5}
\end{figure*}

We finish this section by computing different observables in the states prepared with the hybrid and the fully variational protocols. In Fig.~\ref{fig:4}, we plot different correlation functions, which can be measured in cold-atom experiments using state-of-the-art techniques, such as quantum gas microscopes~\cite{Bakr_2009, Sherson_2010, Gross_2021}. In particular, in Fig.~\ref{fig:4}(a) we show the spin-spin correlations $\langle \hat{S}^{z}_{i} \hat{S}^{z}_{j} \rangle$ between the atoms within one plaquette. We observe how the $\mathrm{HyVA}_{1\rightarrow2}$ protocol does not manage to build the antiferromagnetic correlations present in the exact ground state, as opposed to the fully variational circuit for the same evolution time. In Fig.~\ref{fig:4}(b), we show the spin structure factor,
\begin{equation}
\label{eq:spin-spin-structure-factor}
    S_m (k) = \sum_{i,j} e^{i k (i-j)} \left[\langle \hat{S}_{i}^{z}\hat{S}_{j}^{z}\rangle - \langle\hat{S}_{i}^{z}\rangle\langle\hat{S}_{j}^{z}\rangle\right]\,,
\end{equation}
calculated along the $x-$axis. For this observable, both the fully variational and the $\mathrm{HyVA}_{2\rightarrow3}$ produce similar results, confirming our previous conclusions. Here we note that the extremely good performance of our protocols is partially a consequence of the relatively simple structure of the target ground state. We now demonstrate how these results are maintained for the more complicated case of a FH model with finite doping.

\subsection{Fermi-Hubbard ladders at finite doping~\label{subsec:doping}}

In the presence of a finite doping $\delta$, several works~\cite{BoXiao2017,SimonsCollaboration2020,Jiang2019} have found that the ground state of $\hat{H}_{\mathrm{FH}}$ forms charge stripes with a wavelength $\lambda_e = 1/\delta$, leading to one doped hole per unit cell. We now show how to prepare these doped ground states, which present a smaller energy gap as compared to the half-filled case studied above. In this case, we start with an antiferromagnetic product state of $8$ fermionic atoms on a $2\times 6$ ladder [Fig.~\ref{fig:5}(a)], corresponding to $\delta=1/3$. This state can be obtained experimentally by removing $4$ atoms in non-consecutive rungs of the ladder, starting from the state $\ket{\Psi_0}$ considered above for the half-filled case~\cite{parsons16a,Mazurenko_2017,Koepsell_2019,Hirthe_2023,Hartke_2023, Xu_2025}. 

From this initial state, we prepare the ground state at $U/t=8$ following two steps. First, we prepare the ground state of the dimers at the edge of the system as well as the central $2\times 2$ plaquette, using the pre-compiled variational circuits described above for the half-filled case. We then connect the dimers and plaquettes by applying the following parametrized unitary,
    \begin{align}~\label{eq:U_links_delta}
        \mathcal{U}_{\delta}\left(\bm{\theta}\right) = \prod_{\mu=1}^D e^{-i T_\mu \hat{H}_{\delta}(\bm{\theta}_\mu)}\,,
    \end{align}
    with
    \begin{align}\label{eq:H_FH_links_delta}
        \hat{H}_{\delta}= &-\tilde{t}\sum_{\langle \ii,\jj\rangle \in \mathrm{links},\sigma}\hat{c}_{\ii\sigma}^\dagger \hat{c}_{\jj\sigma}-t\sum_{\langle \ii,\jj\rangle \not\in \mathrm{links},\sigma}\hat{c}_{\ii\sigma}^\dagger \hat{c}_{\jj\sigma}\\
    &+U\sum_\ii \hat{n}_{\ii\uparrow}\hat{n}_{\ii\downarrow}+\Delta/t \sum_{\ii,\sigma}\mu_{\ii}\hat{n}_{\ii,\sigma}\,,
    \end{align}
    and variational parameters $\bm{\theta}_\mu=\left(T_\mu\,\tilde{t}_{\mu},T_\mu\,\Delta_\mu\right)$. Note that $\tilde{t}$ is now the tunneling for the links connecting dimers and plaquettes with their contiguous empty sites, and we fix $\mu_{\ii}/t = 4 $ for the initially empty sites and zero for the half-filled ones. This Hamiltonian can be implemented for larger systems in a translationally-invariant manner while keeping $\delta$ fixed.

In Figs.~\ref{fig:5}(b)--(c), we plot $\varepsilon$ and the $\mathcal{I}$ for the optimized circuit with $D=2$ layers, and compare the results with the ones obtained with a HyVA protocol. For the latter, the adiabatic fusion is now performed in two steps, first by ramping up from $\tilde{t}=0$ to $\tilde{t}=t$, while keeping the $\Delta$ fixed for a time $T_1$, and then decreasing the chemical potentials until they are uniform along the ladder, for a time $T_2$. From these results, we conclude that, similarly to the half-filled case, the fully variational state preparation shows a better performance in terms of both the fidelity and the accuracy of the energy estimation compared to the HyVA method for the same implementation time. However, the advantage of the latter relies on the fact that all the variational circuits are optimized classically, and can be thus applied directly to an extended system. In the insets of the figure, we show how the results obtained with the HyVA protocol can be further improved by increasing the implementation time, which is still within the typical experimental coherence times. In the next section, we show how these results are maintained for larger system sizes.

In Fig.~\ref{fig:5}(d), we plot the charge-density profile along the $x-$axis, $\langle\hat{n}_{x}\rangle$, to show that our variationally prepared states [at the time indicated by the markers in panels (b-c)] reproduce the expected periodicity of the stripe pattern of the exact ground state. In the case of the HyVA protocol, longer implementations times are required for this, which is consistent with the results described above. Finally, the spin-structure factor along the $x-$axis also agree for the states prepared with the different methods, as shown in Fig.~\ref{fig:5}(e). We note that, while the quantitative agreement is not as good as for the half-filled case, our variational protocols approximate the properties of the exact ground state qualitatively for short implementation times. The difference here compared to half filling stems from the smaller gap and therefore longer correlation length of the target ground state. 

\subsection{Extended Fermi-Hubbard ladders~\label{subsec:extendedtun}}

We finish this section by showing how our variational protocols can also be used to prepare the ground states of extended Fermi-Hubbard (eFH) models, where the target and resource Hamiltonians do not match. We illustrate our approach for the eFH model in the presence of NNN tunnelings terms,
\begin{equation}
\hat{H}_\mathrm{eFH}=\hat{H}_\mathrm{FH}+\hat{H}_{t'},
\end{equation}
with
\begin{equation}
   \hat{H}_{t'}=-t'\sum_{\langle\langle\ii,\jj\rangle\rangle,\sigma} \hat{c}_{\ii,\sigma}^\dagger \hat{c}_{\jj,\sigma}\,.~\label{eq:NNN}
\end{equation}
This model has regained significant interest because several works have suggested that these longer-range tunnelings stabilize a $d$-wave superconducting order in the thermodynamic limit~\cite{Jiang2019, Ponsioen_2019, Xu_2024}. However, the question is still not completely settled, since obtaining the ground state for large system sizes using classical methods is computationally very challenging~\cite{Qin_2022}. This and other questions could be further investigated in fermionic cold-atom quantum simulators, where the $d$-wave order parameter can be efficiently measured experimentally~\cite{Mark_2024}.

\begin{figure}[tb]
    \centering
    \includegraphics[width=\linewidth]{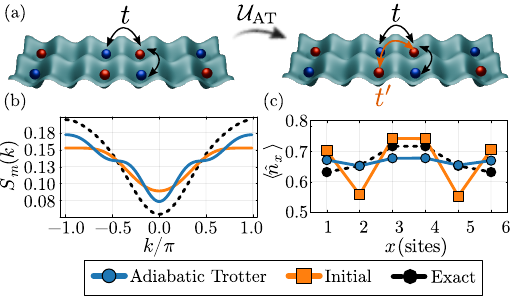}
    \caption{{\bf Extended Fermi-Hubbard model.} (a) Using the $\mathcal{U}_{\mathrm{AT}}$, we prepare the ground state of the eFH model on a $2\times6$ ladder, for $U/t = 8$, $t/t'=-0.25$ and $\delta = 1/3$. As described in the main text, we take as an initial state the ground state of the local FH model prepared with the HyVA protocol (orange) using a total time $T t = 20$, and apply an adiabatic Trotter evolution for a time $T_{\mathrm{Trotter}} t = 10$ to reach the target state (blue). (b) and (c) show the spin structure factor $S_m(k)$ and the atomic density, respectively, calculated along the ladder. We show results for the initial and the prepared state, and compare to the results obtained on the exact ground state of the model (in black).}
    \label{fig:exttun}
\end{figure}

We study in particular the eFH model~\eqref{eq:NNN} on a ladder geometry. We focus on the case with a finite doping of $\delta = 1/3$, and take $U/t = 8$ and $t'/t = -0.25$, a parameter regime that is compatible with the numerical evidence of $d$-wave superconducting order~\cite{Jiang2019}. As initial state for the protocol, we consider the ground state of the local FH Hamiltonian, $\ket{\psi_{\mathrm{FH}}}$, prepared with the protocols introduced in Sec.~\ref{subsec:doping}. We then transform this state to the ground state of the eFH model, $\ket{\psi_{\mathrm{eFH}}}$, through an adiabatic Trotter (AT) evolution, where the NNN tunneling terms are slowly introduced through a linear ramp, as schematically depicted in Fig.~\ref{fig:exttun}(a). In App.~\ref{app:extended_FH}, we show how to generate time evolutions under $\hat{H}_{t'}$ by just concatenating three unitaries from the fermionic gate set $\mathcal{G}$~\eqref{eq:Gateset}, such that the whole protocol can be implemented using only local FH dynamics.

On a ladder geometry, this unitary evolution takes the form
\begin{align}~\label{eq:UAT}
    \mathcal{U}_{\mathrm{AT}} =& \prod_k e^{-i \hat{H}_{\mathrm{FH}}\frac{\Delta T}{2}}e^{-i \hat{H}_{t^\prime}^{1}\frac{\Delta \tilde{T}_k}{2}} \\
    &\times e^{-i \hat{H}_{t^\prime}^{2}\Delta \tilde{T}_k}e^{-i \hat{H}_{t^\prime}^{1}\frac{\Delta \tilde{T}_k}{2}}e^{-i \hat{H}_{\mathrm{FH}}\frac{\Delta T}{2}}\,,\nonumber
\end{align}
with $\Delta\tilde{T}_k = \frac{\left(T_k + T_{k-1}\right)\Delta T}{2 T_{\mathrm{Trotter}}}$, $T_k = k \Delta T $ with $k$ an integer and $\Delta T$ a suitable timestep, and $0 < T_k \leq T_{\mathrm{Trotter}}$. Here $\hat{H}_{t^\prime}^{1(2)}$ refers to the two sets of commuting operators in $\hat{H}_{t^\prime}$, where we choose in particular the parallel diagonal terms along the two possible orientations in the ladder. Finally, we note that here the unitary time evolutions with NNN tunnelings terms, $e^{-i\hat{H}_{t^\prime}\Delta T}$, can be implemented using quench evolutions with only nearest-neighbor tunnelings, as described in App.~\ref{app:extended_FH}, such that the whole protocol can be implemented with the fermionic gate set~\eqref{eq:Gateset}. In this case, the application of the $\mathcal{U}_{\mathrm{AT}}$ operator for a total time $T_{\mathrm{Trotter}}$ needs an equivalent physical time of $T = 2T_{\mathrm{Trotter}}$.

In Fig.~\ref{fig:exttun}, we show the results obtained after applying this AT protocol to the initial state, $\ket{\psi_{\mathrm{FH}}}$, obtained with the HyVA protocol described Sec.~\ref{subsec:doping} for a total implementation time of $T t = 20$. This state corresponds to the ground state of the local FH Hamiltonian prepared with a fidelity of $\sim 87\%$. After its preparation, we implement $\mathcal{U}_{\mathrm{AT}}$ for a total time of $T_{\mathrm{Trotter}} t = 10$ and a timestep $\Delta T t = 0.05$, resulting in a final fidelity of $\sim 86\%$ for the exact ground state of the eFH model. In Figs.~\ref{fig:exttun}(b) and (c), we plot the spin structure factor $S_m(k)$~\eqref{eq:spin-spin-structure-factor} and the occupation numbers along the $x-$axis, respectively, for the initial and the final states after the AT protocol. Our results show how, despite the imperfect state prepared due to Trotter errors arising from the short value of $T_\text{Trottter}$ used here, we can approximate the properties of the ground state of the extended model, that are qualitatively different from those in the initial state.  Specifically, in this case our protocol is able to melt the initial stripe order, as required to achieve a $d$-wave superconducting state. 

The study of the emergence of such order is limited here by the system sizes we can simulate classically. This and other properties of the phase diagram of the  eFH model could be further investigated in a fermionic quantum simulator by applying the state-preparation protocols described here to larger system sizes. Finally, we note that the Trotter circuit can be shortened considerably by promoting the Trotter times to variational parameters, which could be optimized to reach similar fidelities in fewer Trotter steps. While performing this optimization classically is challenging for the system sizes considered here, this could be done directly in the quantum hardware through a VQE protocol.

\section{Variational quantum imaginary-time evolution~\label{sec:ITE}}

In this section, we show how to approximate quantum imaginary-time evolution (QITE) on a fermionic quantum simulator using fermionic unitary circuits. Apart from being an alternative state preparation method, QITE is an important subroutine for other quantum simulation algorithms, such as subspace expansion methods~\cite{Motta_2024_subspace} or the algorithm recently introduced in Ref.~\cite{Yang2024PhaseMeas} to estimate Loschmidt amplitudes. This last algorithm, through the possibility of performing QITE, could be used as a way to implement filtering algorithms~\cite{Poulin2009,Ge2019,Lin_2020,Lu_2021} and statistical phase estimation protocols~\cite{Somma_2019,OBrien_2019,Lin2022,Ding2023QCELS,Wang_2023} with fermionic quantum simulators.

Starting from some initial state $\ket{\Phi(\tau = 0)}$, the QITE protocol prepares the (non-degenerate) ground state of a target Hamiltonian $\hat{H}$,
\begin{equation}
   |\Psi_\mathrm{GS}\rangle = \lim_{\tau\rightarrow\infty} \frac{e^{-\tau \hat{H}}|\Phi(\tau = 0)\rangle}{\lVert e^{-\tau \hat{H}}|\Phi(\tau = 0)\rangle\lVert} \, ,
\end{equation}
provided that these two states have a non-zero overlap, i.e., $\langle \Psi_\mathrm{GS} |\Phi(\tau = 0)\rangle\neq 0$. Implementing this protocol using quantum hardware is challenging since it requires the application of a non-unitary operator. In~\ref{subsec:varQITE}, we introduce a procedure to implement it in a fermionic quantum simulator by adapting a QITE algorithm that was first proposed for qubit-based devices~\cite{motta2020a}, as well as its variational approximation (VarQITE)~\cite{McArdle2019,Yuan2019theoryofvariational,Gacon2024}. In~\ref{subsec:QLanczos}, we explain how to use the VarQITE protocol to construct an approximate Krylov subspace by post-processing the collected data, and then use it to obtain a better estimation of the energy using the Quantum Lanczos (QLanczos) algorithm~\cite{motta2020a}. In~\ref{subsec:resultsQITELancozs}, we illustrate our protocols for the local FH model studied in section~\ref{sec:state}, as well as for an extended FH model with NN interactions. For the latter, the QITE protocol is implemented using only the local fermionic gate set~\eqref{eq:Gateset}, demonstrating that the proposed algorithms can be used to study target models different from the native experimental resources, by performing measurements on the quantum hardware and applying classical post-processing. We finish this section by investigating the effect of shot noise on these algorithms, and giving an estimation on the number of experimental repetitions required to run them with good precision on a cold-atom setup.

\subsection{Approximating QITE with unitary evolutions~\label{subsec:varQITE}}

Given a target Hamiltonian $\hat{H}$, the goal of the protocol is to implement a variational approximation of the non-unitary QITE step,
\begin{equation}~\label{eq:QITE_step}
    \ket{\psi\left(\tau + \Delta\tau\right)} = c_n e^{-\Delta\tau \hat{H}}\ket{\psi\left(\tau\right)}\,,
\end{equation}
where $\Delta\tau$ is the imaginary time step, $\ket{\psi\left(\tau\right)}$ is the imaginary-time evolved state at time $\tau$ obtained after $n$ QITE steps, with $\ket{\psi\left(\tau = 0\right)}$ some initial state, and $c(\tau) = 1/ \lVert e^{-\Delta\tau \hat{H}}\ket{\psi\left(\tau\right)} \rVert$ is a normalization constant. Similar to other variational approaches, the idea is to define a parametrized quantum state $\ket{\psi\left(\bm{\theta}\left(\tau\right)\right)}$, prepared through a variational quantum circuit constructed using quenched unitary dynamics as described in Sec.~\ref{sec:state}, and characterized by a vector of tunable parameters $\bm{\theta}(\tau)\in \mathbb{R}^d$. In Ref.~\cite{Yuan2019theoryofvariational}, the imaginary time dynamics of the parameters at a time $\tau$ are derived from a McLahlan's variational principle. However, as explained in App.~\ref{appendix:ITE}, the required measurements are challenging to obtain with a fermionic optical lattice simulator. For this reason, in App.~\ref{appendix:ITE} we develop an iterative approach similar to the one proposed in Ref.~\cite{motta2020a}, where we approximate instead each time step using the following variational state:
\begin{equation}~\label{eq:parametrization_QITE}
    \ket{\psi\left(\bm{\theta}\left(\tau+\Delta\tau\right)\right)} = \prod_{\mu = 1}^D e^{-i T_\mu \hat{H}_\mu} \ket{\psi\left(\bm{\theta}\left(\tau\right)\right)}\,,
\end{equation}
where $\hat{H}_\mu$ are the different terms of the Hamiltonian, e.g., $\hat{H}_\mu\in\{\hat{H}_t, \hat{H}_U\}$ for the local FH model. The state is thus updated by a sequence of global quenches with variational parameters $\theta_\mu\left(\tau\right) = T_\mu \lambda_{\mu}$, with $\lambda_{\mu} \in \{t, U\}$, corresponding to fermionic gates in $\mathcal{G}$~\eqref{eq:Gateset}. We refer to this approach as a variational QITE (VarQITE). 

Setting the variational parameters to zero at each step, the quantities that need to be measured in the quantum hardware to update the variational parameters are the following,
\begin{align}~\label{eq:objects_to_measure_QITE}
\begin{split}
    g_{\mu\nu} \left(\bm{\theta}\left(\tau\right)\right)  &= \Re\left[\langle \hat{H}_{\mu}\,\hat{H}_{\nu}\rangle_{\bm{\theta}}\right]-\langle\hat{H}_{\mu}\rangle_{\bm{\theta}}\langle\hat{H}_{\nu}\rangle_{\bm{\theta}},\quad \\ b_{\mu} \left(\bm{\theta}\left(\tau\right)\right) &= \Im\left[\langle \hat{H}_{\mu}\,\hat{H}\rangle_{\bm{\theta}}\right]\,,
\end{split}
\end{align}
where $\Re$ and $\Im$ denote the real and imaginary part, respectively. These observables can be measured in a cold-atom setup using state-of-the-art techniques, either globally including extra pulse sequences~\cite{Mark_2024}, or through parallelized measurements with the superlattice as it is explained in App.~\ref{appendix:measurements}. We note that, although the number of terms that we have to measure scale with the system size, making this protocol more challenging that the pre-compiled variational circuits described in the previous section, each of them can be obtained using current experimental techniques~\cite{Impertro_2024}.

\subsection{Using VarQITE to improve energy estimation through QLanczos~\label{subsec:QLanczos}}

Let us now explain how the VarQITE can be used as a subroutine for more complex algorithms. One example is the QLanczos algorithm~\cite{motta2020a}, which can be used to improve significantly the energy estimated in the state prepared by the VarQITE. The key idea of this algorithm is to project the target Hamiltonian $\hat{H}$ into the Krylov subspace spanned by $\ket{\psi\left(\tau_{\alpha}\right)}$,
\begin{equation}
\begin{aligned}
~\label{eq:QLanczos_HandS_full}
\mathbb{H}_{\alpha,\beta}&=\bra{\psi\left(\tau_{\alpha}\right)}\hat{H}\ket{\psi\left(\tau_{\beta}\right)}\,,
\end{aligned}
\end{equation}
at different imaginary times $\tau_\alpha = \alpha\Delta\tau$, with $\alpha = 1,\dots, N$. Using this projection, we can write the effective Schr\"odinger equation in the Krylov subspace, $\mathbb{H}|\Psi\rangle=E\mathbb{S}|\Psi\rangle$, where $\mathbb{S}$ is the matrix that takes into account the normalization,
\begin{equation}
\begin{aligned}
\mathbb{S}_{\alpha,\beta}&=\langle\psi\left(\tau_{\alpha}\right)|\psi\left(\tau_{\beta}\right)\rangle\,.~\label{eq:QLanczos_HandS}
\end{aligned}
\end{equation}

These matrices can be extracted using the quantum simulator in two different ways. First, as it is detailed in App.~\ref{appendix:ITE}, one can measure $|\mathbb{S}_{\alpha,\beta}|$ and $|\mathbb{H}_{\alpha,\beta}|$ directly by reversing the corresponding variational circuits, and then follow the ideas described in Ref.~\cite{Yang2024PhaseMeas} to obtain the corresponding phase. We refer to this approach as the \emph{complete} QLanczos algorithm. Alternatively, one can obtain these matrices approximately using the recursive formula (check App.~\ref{appendix:ITE} for an explicit derivation),
\begin{equation}
\begin{aligned}~\label{eq:matsHandS}
    \mathbb{S}_{\frac{\alpha+1}{2},\frac{\beta+1}{2}} &= c_{\alpha}\,c_\beta /c_{\frac{\alpha+\beta}{2}}^2\,,\\
    \mathbb{H}_{\frac{\alpha+1}{2},\frac{\beta+1}{2}} &= \mathbb{S}_{\frac{\alpha+1}{2},\frac{\beta+1}{2}} \langle\psi(\tau_{\frac{\alpha+\beta}{2}})|\hat{H}|\psi(\tau_{\frac{\alpha+\beta}{2}})\rangle \, ,
\end{aligned}
\end{equation}
where $c_\alpha$ are the normalization constants that can be obtained recursively at each step, by setting $c_1=1$ and
\begin{equation}~\label{eq:normalization_approx}
    c_{\alpha+1}^{2} = \frac{1-2\Delta\tau \bra{\psi\left(\tau_{\alpha}\right)}\hat{H}\ket{\psi\left(\tau_{\alpha}\right)}}{c_{\alpha}^{2}}+\mathcal{O}(\Delta \tau^2)\,.
\end{equation}

The advantage of this approximation, that we denote as \textit{approximated} QLanczos, is that it only relies on the measurement of $\bra{\psi\left(\tau_{\alpha}\right)}\hat{H}\ket{\psi\left(\tau_{\alpha}\right)}$. This magnitude can be obtained by using the collected data from the VarQITE algorithm, avoiding the application of inverse circuits and measurements of Loschmidt amplitudes, as required in the complete QLanczos approach. Let us also mention that even though this approximation relies on a small time step $\Delta\tau$ to reach a good accuracy, the number of measurements required to calculate VarQITE scales with $1/\Delta\tau$, so this time step cannot be made arbitrary small. In App.~\ref{appendix:ITE}, we explain in detail how to measure in a cold-atom setup the required quantities for these two methods.

\subsection{Application of VarQITE and QLanczos to Fermi-Hubbard ladders~\label{subsec:resultsQITELancozs}}

\begin{figure}[tb]
    \centering
    \includegraphics[width=\linewidth]{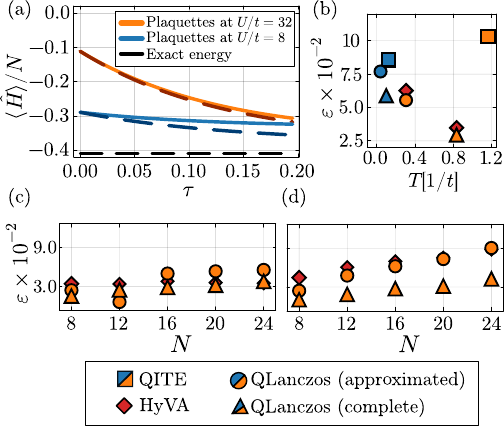}
    \caption{{\bf VarQITE and QLanczos for the local Fermi-Hubbard ladder.} (a) Comparison between the QITE protocol (solid lines) and the exact ITE (dashed lines) to prepare the ground state of a half-filled $2\times12$ ladder at $U/t = 8$, starting from different initial states (see main text). (b) We show the best residual energies $\varepsilon$ obtained with the QITE protocol as well as with the two QLanczos methods, together with the required implementation time $T$, for the different initial states. We also show results obtained with the HyVA method (see Fig.~\ref{fig:3}) using the same implementation time. (c) and (d) show the results obtained for $\varepsilon$ using the different protocols as a function of the ladder length $N$, for the case of half filling and quarter filling, respectively.}
    \label{fig:6}
\end{figure}

We now apply the VarQITE and the different QLanczos protocols discussed above to the Fermi-Hubbard ladder with $U/t=8$, both at half filling and for finite dopings. Moreover, we also consider the effect of NN interaction terms. We investigate the performance of these methods as a function of the system size sizes, and for different choices of the initial state $\ket{\psi(\tau = 0)}$, which can be prepared with the variational protocols described in Sec.~\ref{sec:state}. As shown in Ref.~\cite{motta2020a}, the initial state should have a sufficiently good overlap with the target ground state, while at the same time have a moderate correlation length such that the non-unitary QITE step can be approximated by a unitary operator. For these reasons we choose: i) a tensor product of the ground state of the FH model at $U / t = 8$ for each uncoupled plaquette; ii) a similar plaquette product state, but obtained for $U / t = 32$ to reduce the initial correlation length. 

\subsubsection{Local Fermi-Hubbard ladders}

Our results for the local FH ladder are summarized in Fig.~\ref{fig:6}, where we first show the expectation value of the target Hamiltonian $\mean{\hat{H}}$ over the states evolved with our VarQITE protocol for a $2\times12$ ladder at half filling, and compare them with an exact QITE calculation [black dashed line in Fig.~\ref{fig:6}(a)]. These results show the importance of the choice of the initial state. In particular, if we take the plaquette product state for $U / t = 8$ as the initial state (in solid blue), the protocol does not provide a significant improvement in the energy because the QITE step has a support (i.e. the number of fermion modes in which it acts simultaneously) smaller than the correlation length of the system, and thus it can not properly capture the imaginary time dynamics~\cite{motta2020a}. In the language of variational quantum computing, we would say that the circuit lacks expressibility for that given initial state. On the other hand, for the initial plaquette product state with $U / t = 32$, which has a smaller overlap with the exact ground state, the protocol results in a larger improvement of the initial energy during the optimization. These results exemplify a trade-off between the trainability of the circuit and the quality of the initial state. 

Similarly to the HyVA protocol introduced in the previous section, our results show how our VarQITE approach can prepare the ground state of the FH Hamiltonian much faster than standard adiabatic protocols. As shown in Fig.~\ref{fig:6}(b), the bare application of VarQITE does not provide an advantage over the HyVA approach, considering in particular that it requires more complicated measurements. However, as we mention before, it can be used as a subroutine for other algorithms, such as the QLanczos. We now show how this approach provides an advantage with respect to the HyVA protocol both in terms of implementation times as well as in the accuracy of the estimated ground-state energy.
 
In Fig.~\ref{fig:6}(b)-(d), we summarize our results using the two QLanczos approaches described above. First, in Fig.~\ref{fig:6}(b) we plot the residual energy $\varepsilon$ as a function of the total implementation time $T$, and compare the results obtained directly from the VarQITE algorithm [shown in Fig.~\ref{fig:6}(a)], and the ones obtained from the QLanczos algorithms [circle (triangle) markers correspond to the approximated (complete) method]. We also benchmark these protocols against the HyVA state preparation introduced in the previous section by showing, in each case, results obtained for the same total implementation times [in red rhombus]. From these results, we can first conclude that both QLanczos protocols improve the energy with respect to the VarQITE algorithm from which the information is extracted, and, more importantly, they reduce considerably the evolution time required for that, since only short imaginary times are required. Moreover, for this system size,  QLanczos performs better if the corresponding VarQITE starts with the plaquette state at $U / t = 32$ [orange lines/markers], even if the final energy obtained with the latter is worse than the for initial plaquette state at $U / t = 8$ [blue lines/markers]. The reason is that the former has a shorter correlation length, and approximates better the exact QITE for shorter times. Finally, we also observe how, for this system size, both versions of the QLanczos algorithm introduced above perform better than the HyVA protocol for an equivalent implementation time.

In Fig.~\ref{fig:6}(c), we make a systematic analysis of the QLanczos advantage over the HyVA protocol for increasing ladder sizes, i.e., $N=2\times L$, where $L$ is the length of the ladder. For the half-filled case, we can observe that the approximated QLanczos version shows a worse performance compared to the HyVA protocol after a certain system size $N$. For the complete QLanczos, the performance is similar to the HyVA protocol. This behaviour can be explained by the nature of the state we are trying to prepare. At half filling, the gap of the system is considerably large, and the adiabatic fusion of plaquettes is expected to be efficient even for short implementation times. The situation changes, however, if we now compare the performance of these protocols in the presence of doping. The results are shown Fig.~\ref{fig:6}(d), where we perform a similar scaling analysis of the performance of QLanczos against the HyVA protocol for the case of quarter filling. In that case, we now find a clear advantage of the complete QLanczos compared with the HyVA protocol that is maintained, and even increases, as the system size grows.

\begin{figure}[tb]
    \centering
    \includegraphics[width=\linewidth]{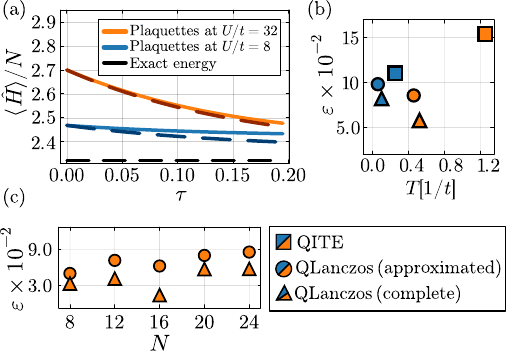}
    \caption{{\bf VarQITE and QLanczos for the Fermi-Hubbard ladder with extended interactions.} (a) Comparison between the QITE protocol (solid lines) and the exact ITE (dashed lines) to prepare the ground state of a half-filled $2\times12$ ladder at $U/t = 8$ and $V/t = 2$, starting from different initial states (see legend). (b) Best residual energies $\varepsilon$ obtained with the QITE protocol as well as with the two QLanczos methods, together with the required implementation time $T$, for the different initial states. (c) Residual energies $\varepsilon$ as a function of the ladder length $N$ for the different protocols.}
    \label{fig:6_extints}
\end{figure}

\subsubsection{Fermi-Hubbard ladder with extended interactions}

One of the main advantages of implementing the VarQITE and the QLanczos protocols is that there is no need to directly engineer all the terms in the target Hamiltonian $\hat{H}$ to estimate its ground-state energy. To illustrate this point, we now include an extra NN interaction term to the half-filled FH ladder target Hamiltonian,
\begin{equation}
    \hat{H}_V = V \sum_{\langle\ii,\jj\rangle}\hat{n}_{\ii}\hat{n}_{\jj}\,,
\end{equation}
where we set $V/t=2$. We apply our protocols to this extended FH ladder, and summarize our results in Fig.~\ref{fig:6_extints}. In this case, both the initial states and the quench evolutions in the fermionic circuits are the same as above. The only difference is that now the observables that we have to measure in~\eqref{eq:objects_to_measure_QITE} also include the new term $\hat{H}_V$, as part of $\hat{H} = \hat{H}_\text{FH} + \hat{H}_V$.

In Fig.~\ref{fig:6_extints}(b), we show the results obtained after implementing the VarQITE protocol starting from the same initial states as in Fig.~\ref{fig:6}(a). There, we demonstrate that we can accurately approximate the exact QITE corresponding to the extended FH Hamiltonian for the chosen parameter regimes. Furthermore, in Figs.~\ref{fig:6_extints}(c-d) we perform a similar system-size analysis as in the local FH case. We find that the QLanczos still provides an advantage both in its approximate and its complete form when compared to the regular VarQITE, and that these results hold as the system size increases. We note that, in this case, we do not assume that the term $\hat{H}_V$ is generated on the quantum simulator. Therefore, and since there is no equivalent adiabatic preparation, our results show how both the VarQITE and the QLanczos protocols developed here can be used to explore target models beyond the ones engineered in the experiment.

\begin{figure}[tb]
    \centering
    \includegraphics[width=\linewidth]{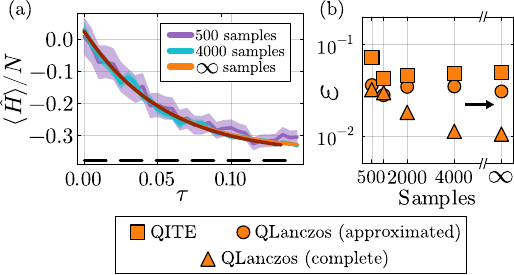}
    \caption{{\bf Effects of shot noise in Quantum Imaginary Time Evolution (QITE) and QLanczos.} (a) Comparison between the QITE (solid lines) and the exact imaginary time Evolution (dashed lines) applied to the target local FH model with $U/t = 8$ in a $2\times4$ ladder, considering as initial state the uncoupled plaquettes with reduced $t$ in Fig.~\ref{fig:6}(a). Each line considers a different number of samples per object, with the shaded ribbon being the standard deviation. (b) Best residual energies $\varepsilon$ compared with the number of samples per object used to obtain them when considering the situation depicted in panel (a).}
    \label{fig:6_2}
\end{figure}

\subsubsection{Effect of shot noise}

Finally, let us note that all the results shown for the VarQITE and the QLanczos algorithm so far assume an infinite number of measurements. We now conclude this study by investigating how our protocols are affected by the presence of shot noise stemming from a finite number of samples in the estimation of each measured object. In Fig.~\ref{fig:6_2}(a), we compare the results obtained from the VarQITE algorithm for different number of samples, showing how the protocol is still able to capture the results obtained using an infinite number of measurements. Then, in Fig.~\ref{fig:6_2}(b), we show the final estimation of the energy provided by the approximated and complete versions of QLanczos as a function of the number of samples. The main conclusion is that the QLanczos provides a good estimation of the energy of the ground state while requiring less coherence time than other methods, even in the presence of shot noise.

The number of required samples can be connected to the number of experimental runs needed to implement the algorithms in practice. Assuming that the required observables are measured following the parallelization schemes proposed in App.~\ref{appendix:measurements}, the number of experimental runs needed to implement a single VarQITE step grow as $\mathcal{O}(8NM)$ for a ladder, with $N$ being the total number of sites and $M$ the number of samples assumed in the calculation [$\mathcal{O}(10NM)$ for a full lattice]. Assuming the possibility to implement in parallel $\sim 10$ independent 2D experiments using a 3D optical lattice, the results shown above for a $2\times4$ ladder and  $M=2000$ samples per object can be obtained with approximately $ 1.5\times10^4$ total experimental runs. This is within reach considering the repetition rates of current cold-atom experiments~\cite{chalopin2024probing,su2025fast}. We note that this number can be further reduced by implementing more sophisticated measurement schemes based on quantum control techniques~\cite{Mark_2024} or randomized protocols~\cite{Naldesi2024_randomized,Gluza2021_correlations,Zhao2021_fermionictomography,low2024classicalshadowsfermionsparticle,denzler2023learningfermioniccorrelationsevolving,Tran2023_measuringanalog}.

\section{Conclusions \& outlook~\label{sec:conclusion}}

In summary, we have shown how cold-atoms in optical lattices can be operated as programmable quantum simulators of fermionic models, where we harness the latest experimental advances demonstrating a local control of these platforms. Here we focus on ground-state preparation, and develop quantum algorithms that overcome some of the limitations of more traditional analogue quantum simulation approaches. 

We first construct hybrid variational-adiabatic protocols that are pre-compiled classically, and demonstrate they can faithfully prepare the ground state of translational-invariant local and extended FH models, both at half filling as well as for finite dopings. Importantly, we show these pre-compiled circuits require considerably less time than adiabatic protocols. Besides, we also demonstrate they can be used to study models beyond the local FH models used as resources, e.g., FH with extended tunneling terms, required to obtain a $d$-wave superconducting order. Second, we propose how to obtain a variational approximation of the quantum imaginary time evolution using resources available in cold-atom setups, and benchmark its potential as an alternative state-preparation strategy for fermionic Hamiltonians. Finally, we also show this protocol can be used as a subroutine for the Quantum Lanczos algorithm, which only requires post-processing the data measured during imaginary time evolution, and demonstrate how it considerably improves the ground-state energy estimation for local and extended FH models. While we benchmark our protocols on ladder geometries, they are designed for extended 2D systems, and can be readily applied to cold-atom experiments to investigate system sizes that can not be addressed using classical methods.

The fermionic gate set considered in this work can be extended by including more general resources that can be natively implemented in cold-atom simulators, such as synthetic gauge fields~\cite{Aidelsburger_2018} or long-range interactions~\cite{Chomaz_2023, Arguello_Luengo_2022}. The former would allow to target fractional quantum Hall states~\cite{Leonard_2023}, which are topologically-ordered phases that are very hard to prepare adiabatically due to gap closing transitions, and the latter could help to prepare critical states more efficiently~\cite{Tabares_2023}.

Besides these specific applications to ground-state preparation, our programmable quantum simulation approach can be used to design and implement other quantum algorithms tailored to cold-atom experiments. The fermionic gate set~\eqref{eq:Gateset} introduced here can be readily used to implement a Trotter real-time evolution under extended FH dynamics. Furthermore, these resources can be also employed to construct more efficient many-body spectroscopy methods to better estimate the properties of the fermionic states prepared here, a research direction that we will explore in future works.

\acknowledgements

D.G.-C. acknowledges support from the European Union's Horizon Europe program under the Marie Sk{\l}odowska Curie Action PROGRAM (Grant No. 101150724). C.T. acknowledges support from Comunidad
de Madrid (PIPF-2022/TEC-25625) and also from Fundaci\'on
Humanismo y Ciencia. C.T. and A.G.T. acknowledge support from the Proyecto Sin\'ergico CAM 2020 Y2020/TCS-6545 (NanoQuCo-CM), the CSIC Research Platform on Quantum Technologies PTI-001 and from Spanish projects PID2021-127968NB-I00 funded by MICIU/AEI/10.13039/501100011033/ and by FEDER Una manera de hacer Europa, TED2021-130552B-C22 funded by  MICIU/AEI /10.13039/501100011033 and by the European Union NextGenerationEU/ PRTR, respectively, and from the QUANTERA project MOLAR with reference PCI2024-153449 and funded MICIU/AEI/10.13039/501100011033 and by the European Union. This research was supported in part by grant no. NSF PHY-2309135 to the Kavli Institute for Theoretical Physics (KITP). The numerical calculations in this paper were performed using the ITensor package~\cite{itensor} and were run on the FinisTerrae supercomputer supported by the Centro de Supercomputaci\'on de Galicia (CESGA). Work in Innsbruck was supported by the European Union's Horizon Europe programmes HORIZON-CL4-2022-QUANTUM-02-SGA via the project 101113690 (PASQuanS2.1).

\appendix


\section{Programming extended Fermi-Hubbard models\label{app:extended_FH}}

\begin{figure}[tb]
    \centering

   \includegraphics[width=\linewidth]{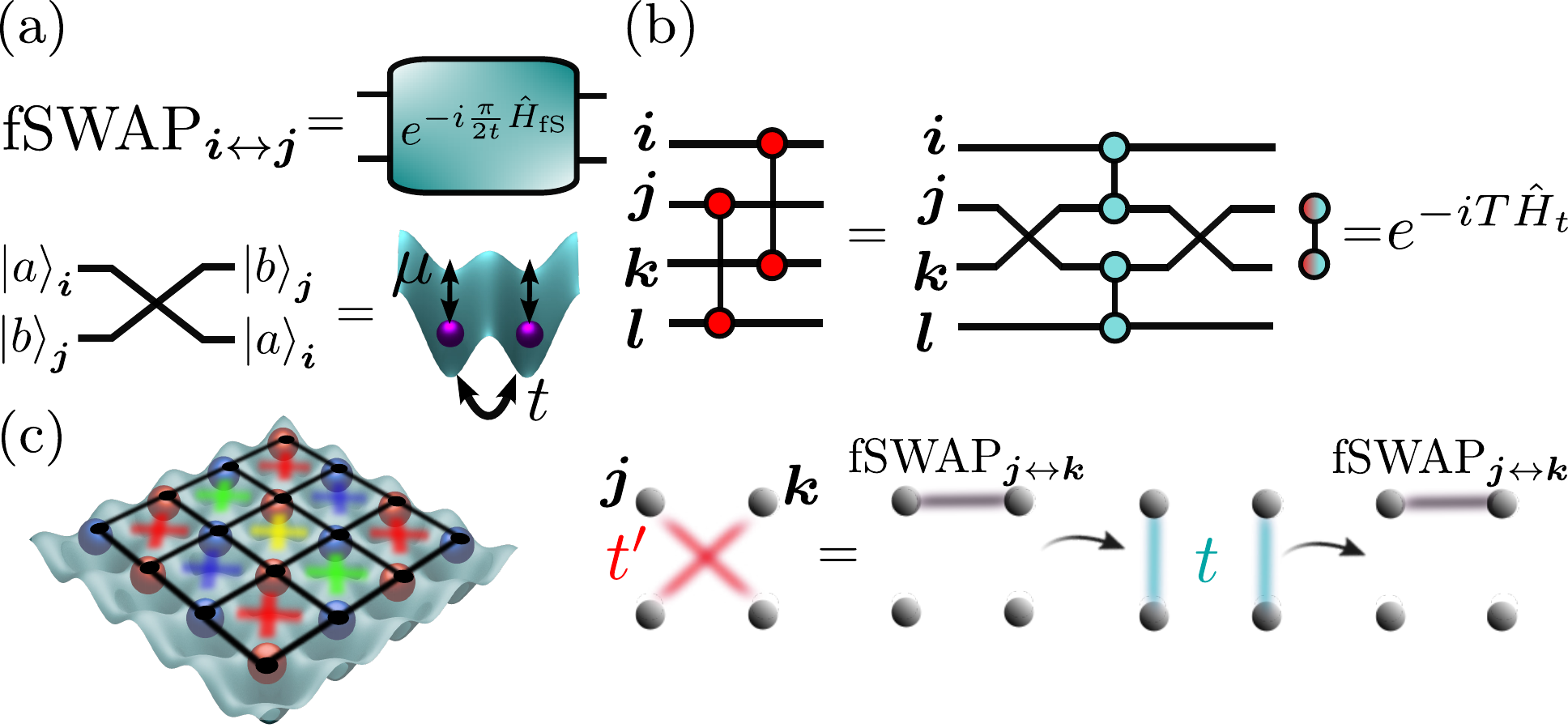}
    \caption{{\bf Fermionic circuits for extended tunneling,} (a) a fSWAP gate can be implemented between two NN sites $\ii$ and $\jj$ by evolving with the Hamiltonian $\hat{H}_{\mathrm{fS}}$~\eqref{eq:FSH} for a time $\pi/(2t)$. (b) A unitary evolution under NNN tunneling terms $\hat{H}_{t^\prime}$ (red, diagonal in the plaquette) can be implemented using a sequence of fSWAP gates, and quench evolutions under the NN tunneling Hamiltonian $\hat{H}_t$ (blue) for a time $tT = t^\prime T^\prime$.
    (c) A sequential application of the operations described in (b) to the four sets of non-commuting plaquettes generates an effective evolutions under NNN neighbor tunneling terms along the whole lattice. 
    }
    \label{fig:2_2}
\end{figure}

In this Appendix, we show how, by implementing a sequence of quench evolutions corresponding to the fermionic gate set $\mathcal{G}$~\eqref{eq:Gateset}, it is possible to generate evolutions under more complicated Hamiltonians beyond the local FH model~\eqref{eq:FH}. While here we focus on the implementation of NNN tunneling term of~\eqref{eq:NNN}, other extended terms can be also obtained in a similar manner.

One key element to achieve this task is the implementation of fermionic SWAP (fSWAP) gates between two nearest-neighboring sites $\ii$ and $\jj$, i.e., $\mathrm{fSWAP}_{\ii\leftrightarrow\jj}\ket{a}_\ii \ket{b}_\jj=\ket{b}_\ii \ket{a}_\jj$, where $\ket{a}$ and $\ket{b}$ are the different atomic states, with $a,\,b\in\{0,\uparrow,\downarrow, \uparrow\downarrow\}$. These gates can be directly implemented by dimerizing the lattice, as described in Fig.~\ref{fig:1}(a), and applying an off-site chemical potential such that the system evolves under the Hamiltonian
\begin{align}
 \hat{H}_{\mathrm{fS}} = -t\sum_\sigma  \left(\hat{c}_{\ii,\sigma}^{\dagger}\hat{c}^{\vphantom{\dagger}}_{\jj,\sigma} + \text{H.c.}\right)-3t\left(\hat{n}_{\ii,\sigma} + \hat{n}_{\jj,\sigma}\right)\,,\label{eq:FSH}
\end{align}
for a total time $T_{\mathrm{fS}}=\pi/(2t)$~\cite{Stanisic2022}, see Fig.~\ref{fig:2_2}(a). Using the superlattice structures shown in Fig.~\ref{fig:1}(a), this gate can be parallelized for all dimers in the lattice, and alternated in different directions to generate more complex evolutions. 

In Fig.~\ref{fig:2_2}(b), we schematically depict how to obtain unitary evolutions under the NNN tunneling Hamiltonian in a plaquette through a sequence of fSWAP gates and dimerized tunnelings. Specifically,
\begin{align}~\label{eq:effective_HNNN}
 e^{-i\hat{H}_{t'}T^\prime} = \mathrm{fSWAP}_{\bm{j}\leftrightarrow \bm{k}} \,e^{-i\hat{H}^{(1)}T_1} \, \mathrm{fSWAP}_{\bm{j}\leftrightarrow \bm{k}}\,,
\end{align}
where $e^{-i\hat{H}^{(1)}T}$ corresponds to a quench evolution for a lattice configuration with vertical tunnelings [see Fig.~\ref{fig:1}(a)], and $tT = t^\prime T^\prime$. This protocol can be applied in parallel to all commuting plaquettes in the lattice [see Fig.~\ref{fig:2_2}(c)], and thus, by repeating it four times, we can implement the unitary evolution under the NNN tunneling Hamiltonian on the whole lattice. We note that this circuit has a finite depth, independent on the system size.

\section{Resource Hamiltonians for pre-compiled variational circuits\label{app:Hamiltonians}}

Here, we write down explicitly the resource Hamiltonians used in Sec.~\ref{sec:state} to prepare the ground state of the FH model at half filling. For the first step of the protocol, the variational circuit~\eqref{eq:U_circuit} is obtained by quenching with the following Hamiltonian,
\begin{equation}
\begin{aligned}
~\label{eq:H_mu_dimer}
    \hat{H}^{(1)} = &-t\sum_{\langle \ii,\jj\rangle\in\mathrm{dimers},\sigma} \hat{c}_{\ii\sigma}^\dagger \hat{c}_{\jj\sigma}+U\sum_\ii \hat{n}_{\ii\uparrow}\hat{n}_{\ii\downarrow}\\
    &-\mu\sum_{\ii}\left(-1\right)^{i_x + i_y}\left(\hat{n}_{\ii,\uparrow}-\hat{n}_{\ii,\downarrow}\right)\,\,,
\end{aligned}
\end{equation}
where the tunneling is non-zero only between pairs of sites within dimers, as depicted in Fig.~\ref{fig:3}(b), and we optimize $t$, $U$ and $\mu$ for each quench.

In the second step, we start with the dimer product state prepared variationally, and construct a second variational circuit by quenching with the Hamiltonian [Fig.~\ref{fig:3}(c)]
\begin{equation}
\begin{aligned}
~\label{eq:H_links}
    &\hat{H}^{(2)}= -\tilde{t}\sum_{\langle \ii,\jj\rangle \in \mathrm{rungs},\sigma}\hat{c}_{\ii\sigma}^\dagger \hat{c}_{\jj\sigma}\\
    &-t\sum_{\langle \ii,\jj\rangle\in \mathrm{dimers},\sigma}\hat{c}_{\ii\sigma}^\dagger \hat{c}_{\jj\sigma}+U\sum_\ii \hat{n}_{\ii\uparrow}\hat{n}_{\ii\downarrow}\,,
\end{aligned}
\end{equation}
where in this case we fix $U/t=8$, and consider only $ \tilde{t}_\mu$ and the quench time $T^{(2)}_\mu$ as variational parameters. Finally, in the third step we apply the Hamiltonian $H^{(3)}$, which is similar to~\eqref{eq:H_links}, but with $\tilde{t}$ and $t$ referring now to the inter-plaquette and intra-plaquette couplings, respectively.

\section{Measurement schemes}~\label{appendix:measurements}

In this Appendix, we show in detail how to measure $\langle \hat{H} \rangle$ for the local FH Hamiltonian, where $\mean{\hat{H}} = \mean{\hat{H}_t} + \mean{\hat{H}_U}$. Extended tunneling terms can be included in a similar manner, by bringing them first to a NN form using a constant-depth circuit of fSWAP gates, as explained in App.~\ref{app:extended_FH}. We also consider measurements of $\mean{\hat{H}_\mu \hat{H}_\nu}$, which are required to implement the quantum imaginary-time evolution. Both $\mean{\hat{H}_U}$ and $\mean{\hat{H}^2_U}$ are diagonal in the occupation basis, and can be measured in a straightforward manner using a quantum gas microscope~\cite{Bakr_2009, Sherson_2010, Gross_2021}. We now consider the remaining terms separately. Finally, let us also note that these measurements can also be performed using the protocol introduced in Ref.~\cite{Mark_2024} and based on quantum optimal control. 

\subsection{Measurement of $\langle\hat{H}_t\rangle$}

The first operator that we are interested in measuring is $\hat{H}_t$. When computed over a two-dimensional square lattice, this operator can be divided in four commuting parts such as all the terms in each one of them can be measured simultaneously through the appropriate dimerization of the lattice. Labeling these commuting parts with the index $\alpha$ yields the decomposition
\begin{equation}~\label{eq_appendix:H_tk}
    \hat{H}_t=-t\sum_\alpha \hat{H}_{t,\alpha} =-t\sum_\alpha \sum_{\langle \ii_\alpha,\jj_\alpha\rangle,\sigma} \hat{c}_{\ii_\alpha,\sigma}^{\dagger} \hat{c}_{\jj_\alpha,\sigma}\,,
\end{equation}
where all the terms in each $\hat{H}_{t,\alpha}$ for different $\ii_\alpha$ and $\jj_\alpha$ commute. The measurement of each term $\hat{H}_{t,\alpha}$ can be performed by first applying the unitary operation
\begin{equation}
    \mathcal{B}_\alpha = \bigotimes_{\langle \ii_\alpha ,\jj_\alpha\rangle,\sigma} \mathcal{B}_{\ii_\alpha,\jj_\alpha,\sigma}\,.
\end{equation}
where each term $\mathcal{B}_{\ii_\alpha,\jj_\alpha,\sigma}$ brings the corresponding tunneling term to a diagonal form in the occupation basis, and they can all be applied in parallel for a fixed $\alpha$, i.e.,
\begin{align}~\label{eq_appendix:Htk_meas}
\begin{split}
    \langle \hat{H}_{t,\alpha}\rangle &= 
    -\frac{t}{2}\sum_{\langle\ii_\alpha , \jj_\alpha \rangle,\sigma }\langle \hat{c}_{\ii_\alpha,\sigma}^{\dagger}\hat{c}_{\jj_\alpha,\sigma} + \hat{c}_{\jj_\alpha,\sigma}^{\dagger}\hat{c}_{\ii_\alpha,\sigma}\rangle_{\psi} \\
    &= -\frac{t}{2} \sum_{\langle\ii_\alpha , \jj_\alpha \rangle,\sigma } \langle \hat{n}_{\ii_\alpha,\sigma}\rangle_\varphi - \langle \hat{n}_{\jj_\alpha,\sigma}\rangle_\varphi\,,
\end{split}
\end{align}
where $\ket{\varphi} = \mathcal{B}_\alpha\ket{\psi}$. For a fixed $\alpha$, all the terms $\langle \hat{H}_{t,\alpha}\rangle$ can thus be measured simultaneously using a quantum gas microscope, as shown experimentally in Ref.~\cite{Impertro_2024}. We note that in~\eqref{eq_appendix:Htk_meas} we have symmetrized each of the NN terms, and thus the addition of the factor $1/2$.

Let us finally show that the $\mathcal{B}$ operator can be alternatively implemented in a compact manner by using two quench evolutions under the Hamiltonian
\begin{align}
\hat{H}_{\mathrm{FT}}\left(\bm{\theta}_{k}\right) = -&\sum_{\sigma}\theta_{\ii\jj}\left(\hat{c}_{\ii,\sigma}^{\dagger}\hat{c}_{\jj,\sigma} + \hat{c}_{\jj,\sigma}^{\dagger}\hat{c}_{\ii,\sigma}\right)\\
-&\mu_{\ii\jj,1}\hat{n}_{\ii,\sigma} -\mu_{\ii\jj,2}\hat{n}_{\jj,\sigma}\,,\nonumber
\end{align}
where $\theta_{\ii\jj}$, $\mu_{\ii\jj,1}$ and $\mu_{\ii\jj,2}$ are different variational parameters for each quench. We can optimize these parameters to minimize the infidelity $\mathcal{I}\left(\bm{\theta}\right) = ||\mathcal{B}-e^{-i\hat{H}_{\mathrm{FT}}\left(\bm{\theta}_1\right)}e^{-i\hat{H}_{\mathrm{FT}}\left(\bm{\theta}_2\right)}||$, where $||.||$ is the Frobenius matrix norm. This protocol results in $\mathcal{I}<10^{-8}$ for a wide range of parameters, that can potentially lead to a faster implementation compared to the $Z_{\pi/2}$ and $X_{\pi/2}$ rotations implemented in Ref.~\cite{Impertro_2024}.

\subsection{Measurement of $\langle\hat{H}^2_t\rangle$}

Now we turn our attention to the measurement of $\langle\hat{H}^2_t\rangle$, corresponding to a linear combination of $O(N^2)$ four point correlation functions
\begin{align}~\label{eq_appendix:four_terms_Ht2}
\begin{split}
    &\langle\hat{H}^2_t\rangle = \sum_{\langle \ii,\jj\rangle,\sigma}\sum_{\langle\bm{r},\bm{s}\rangle,\sigma'}\langle \hat{c}^{\dagger}_{\ii,\sigma} \hat{c}_{\jj,\sigma}\hat{c}^{\dagger}_{\bm{r},\sigma'} \hat{c}_{\bm{s},\sigma'}\rangle\\
    &= \frac{1}{4}\sum_{\langle \ii,\jj\rangle,\sigma}\sum_{\langle\bm{r},\bm{s}\rangle,\sigma'}\left\langle\left( \hat{c}^{\dagger}_{\ii,\sigma} \hat{c}_{\jj,\sigma}+ \mathrm{H.c.}\right)\left(\hat{c}^{\dagger}_{\bm{r},\sigma'} \hat{c}_{\bm{s},\sigma'} +\mathrm{H.c.}  \right)\right\rangle\,.
\end{split}
\end{align}
We now introduce a measurement scheme for these terms and how it can be parallelized.

Let us first consider the case where the following conditions are satisfied: 1) $\ii$ and $\jj$ are both different of $\bm{r}$ and $\bm{s}$, 2) $\sigma\neq \sigma'$, or 3) $\ii = \rr$, $\jj = \ss$ \emph{and} $\sigma = \sigma'$. In that case, it is possible to independently apply an operator $\mathcal{B}_{\ii,\jj}\otimes \mathcal{B}_{\bm{r},\bm{s}}$ that simultaneously diagonalizes each term in parenthesis in~\eqref{eq_appendix:four_terms_Ht2}. Let $\ket{\varphi} = \mathcal{B}_{\ii,\jj}\otimes \mathcal{B}_{\bm{r},\bm{s}}\ket{\psi}$. Then:
    \begin{align}
        \begin{split}
            &\left\langle\left( \hat{c}^{\dagger}_{\ii,\sigma} \hat{c}_{\jj,\sigma}+ \mathrm{H.c.}\right)\left(\hat{c}^{\dagger}_{\bm{r},\sigma'} \hat{c}_{\bm{s},\sigma'} +\mathrm{H.c.}  \right)\right\rangle_\psi = \\
            &=\Big\langle\Big(\hat{n}_{\ii,\sigma}-\hat{n}_{\jj,\sigma}\Big)\Big(\hat{n}_{\bm{r},\sigma'}-\hat{n}_{\bm{s},\sigma'}\Big)\Big\rangle_{\varphi}\,,
        \end{split}
    \end{align}
    and the correlation functions can be computed with measurements over the occupation basis. Furthermore, $O(N)$ of these measurements can be parallelized, so the total number of configurations that need to be measured are of order $O(N)$.

There are still $O(N)$ observables that remain to be measured, namely, when $\ii$ or $\jj$ equal $\bm{r}$ or $\bm{s}$ (not both) \emph{and} $\sigma = \sigma'$. Let $\jj=\bm{r}$ without loss of generality. In that case, each of the terms to measure~\eqref{eq_appendix:four_terms_Ht2} can be rearranged as
    \begin{widetext}
    \begin{align}~\label{eq_appendix:off_diag_Ht2}
        \begin{split}
            \left(\hat{c}^{\dagger}_{\ii,\sigma} \hat{c}_{\jj,\sigma} + \hat{c}^{\dagger}_{\jj,\sigma} \hat{c}_{\bm{s},\sigma} + \mathrm{H.c.}\right)^2 &= \Big(\hat{c}^{\dagger}_{\ii,\sigma} \hat{c}_{\jj,\sigma}+\mathrm{H.c.}\Big)^2 + \Big(\hat{c}^{\dagger}_{\jj,\sigma} \hat{c}_{\bm{s},\sigma}+\mathrm{H.c.}\Big)^2 \\
            &+ \Big(\hat{c}^{\dagger}_{\jj,\sigma} \hat{c}_{\bm{s},\sigma}+\mathrm{H.c.}\Big)\Big(\hat{c}^{\dagger}_{\ii,\sigma} \hat{c}_{\jj,\sigma}+\mathrm{H.c.}\Big) + \Big(\hat{c}^{\dagger}_{\ii,\sigma} \hat{c}_{\jj,\sigma}+\mathrm{H.c.}\Big)\Big(\hat{c}^{\dagger}_{\jj,\sigma} \hat{c}_{\bm{s},\sigma}+\mathrm{H.c.}\Big)\,.
        \end{split}
    \end{align}
    \end{widetext}
    The term $\mathcal{D}_{\ii\jj\bm{s}\sigma} = \Big(\hat{c}^{\dagger}_{\ii,\sigma} \hat{c}_{\jj,\sigma}+\mathrm{H.c.}\Big)^2 + \Big(\hat{c}^{\dagger}_{\jj,\sigma} \hat{c}_{\bm{s},\sigma}+\mathrm{H.c.}\Big)^2$ is already measured in the different permutations considered above, so the only task left is the measurement of the second line in~\eqref{eq_appendix:off_diag_Ht2}. To do so, one may notice that $\hat{c}^{\dagger}_{\jj,\sigma} \hat{c}_{\bm{s},\sigma}\hat{c}^{\dagger}_{\jj,\sigma} \hat{c}_{\ii,\sigma}$, $\hat{c}^{\dagger}_{\bm{s},\sigma} \hat{c}_{\jj,\sigma}\hat{c}^{\dagger}_{\ii,\sigma} \hat{c}_{\jj,\sigma}$ and their hermitian conjugates all are equal to zero due to the fermionic anticommutation rules. Hence,~\eqref{eq_appendix:off_diag_Ht2} reads:
    \begin{align}~\label{eq_appendix:off_diagonal_Ht2_final}
        \begin{split}
            &\left(\hat{c}^{\dagger}_{\ii,\sigma} \hat{c}_{\jj,\sigma} + \hat{c}^{\dagger}_{\jj,\sigma} \hat{c}_{\bm{s},\sigma} + \mathrm{H.c.}\right)^2 = \mathcal{D}_{\ii\jj\bm{s}\sigma}+\\
            &+ \left(\hat{c}_{\ii,\sigma}^\dagger \hat{c}_{\jj,\sigma}\hat{c}_{\jj,\sigma}^\dagger \hat{c}_{\bm{s},\sigma}+ \hat{c}_{\jj,\sigma}^\dagger \hat{c}_{\ii,\sigma}\hat{c}_{\bm{s},\sigma}^\dagger \hat{c}_{\jj,\sigma}+\mathrm{H.c.}\right)\\
            &= \mathcal{D}_{\ii\jj\bm{s}\sigma} + \left(\hat{c}_{\ii,\sigma}^{\dagger}\hat{c}_{\bm{s},\sigma}+\mathrm{H.c.}\right)\Big(1-2\hat{n}_{\jj,\sigma}\Big)\,,
        \end{split}
    \end{align}
    where the final equality has been obtained after an application of the fermionic anticommutation rules. Thus, the procedure to measure~\eqref{eq_appendix:off_diagonal_Ht2_final} consists in the simultaneous measurement of $\left(\hat{c}_{\ii,\sigma}^{\dagger}\hat{c}_{\bm{s},\sigma}+\mathrm{H.c.}\right)$ and $\hat{n}_{\jj,\sigma}$, which can be achieved applying a $\mathrm{fSWAP}_{\bm{s}\leftrightarrow\jj}$ and then applying the corresponding $\mathcal{B}$ operator. Once again, most of these measurements can be parallelized, so it is possible to measure these terms in constant depth.

\subsection{Measurement of $\Re\left[\left\langle\hat{H}_t \hat{H}_U\right\rangle\right]$ and $\Im\left[\left\langle\hat{H}_t \hat{H}_U\right\rangle\right]$}

Finally, let us outline the measurement of the crossed terms, which are necessary for the QITE protocol in Sec.~\ref{sec:ITE}. For the first case, let us write $\Re\left[\left\langle\hat{H}_t \hat{H}_U\right\rangle\right]$ as
\begin{equation}~\label{eq_appendix:crossed_term_real}
    \Re\left[\left\langle\hat{H}_t \hat{H}_U\right\rangle\right] = \frac{1}{2}\left( \left\langle\hat{H}_t \hat{H}_U+\hat{H}_U \hat{H}_t\right\rangle\right)\,.
\end{equation}
Since $\hat{H}_t$ and $\hat{H}_U$ are both Hermitian operators, the second equality in~\eqref{eq_appendix:crossed_term_real} is the expectation value of an Hermitian operator. Decomposing these operators in observables that can be measured, we find that:
\begin{align}~\label{eq_appendix:cross_terms_sum}
    \begin{split}
    &\left\langle\hat{H}_t \hat{H}_U+\hat{H}_U \hat{H}_t\right\rangle =\\&= \frac{1}{2}\sum_{\langle\ii,\jj\rangle,\sigma}\sum_{\bm{s}}\left\langle\left(\hat{c}_{\ii,\sigma}^{\dagger}\hat{c}_{\jj,\sigma}+\hat{c}_{\jj,\sigma}^{\dagger}\hat{c}_{\ii,\sigma}\right)\hat{n}_{\bm{s},\uparrow}\hat{n}_{\bm{s},\downarrow}+\mathrm{H.c.}\right\rangle
    \end{split}
\end{align}

There are two different types of terms arising from the sum above. First, the ones obtained when $\ii,\jj\neq\bm{s}$. In such case, it is possible to diagonalize the $\hat{c}_{\ii,\sigma}^{\dagger}\hat{c}_{\jj,\sigma}+\hat{c}_{\jj,\sigma}^{\dagger}\hat{c}_{\ii,\sigma}$ term using a $\mathcal{B}_{\ii,\jj,\sigma}$ operator and then compute the expectation value as described in the cases above, measuring the products of number operators in the occupation basis. The measurement of these terms can be parallelized so only $O(N)$ different configurations are needed to measure the $O(N^2)$ terms.

Second, there are also the terms where $\ii$ or $\jj$ equal $\bm{s}$. Without loss of generality, let $\jj=\bm{s}$. Then, these terms in~\eqref{eq_appendix:cross_terms_sum} become:
    \begin{align}
        \begin{split}
            &\left\langle\left(\hat{c}_{\ii,\sigma}^{\dagger}\hat{c}_{\jj,\sigma}+\hat{c}_{\jj,\sigma}^{\dagger}\hat{c}_{\ii,\sigma}\right)\hat{n}_{\jj,\uparrow}\hat{n}_{\jj,\downarrow}+\mathrm{H.c.}\right\rangle =\\
            &= \left\langle\hat{c}_{\ii,\sigma}^{\dagger}\hat{c}_{\jj,\sigma}\hat{n}_{\jj,\uparrow}\hat{n}_{\jj,\downarrow}+\mathrm{H.c.}\right\rangle \\
            &= \left\langle\left(\hat{c}_{\ii,\sigma}^{\dagger}\hat{c}_{\jj,\sigma}+\mathrm{H.c.}\right)\hat{n}_{\jj,\sigma'\neq\sigma}\right\rangle\,.
        \end{split}
    \end{align}
    As in the previous cases, the tunneling term can now be diagonalized and hence measured.

Finally, regarding the measurement of $\Im\left[\left\langle\hat{H}_t \hat{H}_U\right\rangle\right]$, we first rewrite this term as
\begin{equation}~\label{eq_appendix:crossed_term_imag}
    \Im\left[\left\langle\hat{H}_t \hat{H}_U\right\rangle\right] = -\frac{i}{2}\left( \left\langle\hat{H}_t \hat{H}_U-\hat{H}_U \hat{H}_t\right\rangle\right)\,.
\end{equation}
The operator to measure in the second equality of~\eqref{eq_appendix:crossed_term_imag} is Hermitian, and hence an observable. Its measurement is completely equivalent to the measurement of $\Re\left[\left\langle\hat{H}_t \hat{H}_U\right\rangle\right]$ that was described above, with the only difference now that the measurement of each of the tunneling terms $\left(\hat{c}_{\ii,\sigma}^{\dagger}\hat{c}_{\jj,\sigma}+\hat{c}_{\jj,\sigma}^{\dagger}\hat{c}_{\ii,\sigma}\right)$ is now a measurement of the current operator $i\left(\hat{c}_{\ii,\sigma}^{\dagger}\hat{c}_{\jj,\sigma}-\hat{c}_{\jj,\sigma}^{\dagger}\hat{c}_{\ii,\sigma}\right)$. Fortunately, such operator can also be diagonalized with a $\mathcal{C}$ operator similar to the $\mathcal{B}$ that diagonalized the tunneling term. As it is described in Ref.~\cite{Impertro_2024}, this term can be easily obtained with a simple $X_{\pi/2}$ rotation.

\section{Implementing Quantum Imaginary Time Evolution in a programmable simulator~\label{appendix:ITE}}

The quantum imaginary-time evolution (QITE) approximately solves the Wick-rotated Schr\"{o}dinger equation in imaginary time $\tau \equiv iT$ to find an approximation of the ground state of a Hamiltonian $\hat{H}$. The equation
\begin{equation}~\label{eq:appendix_qiteeq}
    \frac{\partial}{\partial\tau} |\Psi(\tau)\rangle = -\left(\hat{H}-E_\tau\right) |\Psi(\tau)\rangle\,,
\end{equation}
with $E_\tau = \braket{\psi(\tau)|\hat{H}|\psi(\tau}$, is formally solved for an initial state $\ket{\Psi_0}$ as
\begin{equation}
    \ket{\Psi(\tau)} = \frac{e^{-\tau\hat{H}}\ket{\Psi_0}}{\sqrt{\braket{\Psi_0|e^{-2\tau\hat{H}}|\Psi_0}}}\,.
\end{equation}
In the eigenbasis of $\hat{H}$, $\{\ket{\varphi_i}\}$, the solution $\ket{\Psi(\tau)}$ reads
\begin{equation}~\label{eq:appendix_solqite}
    \ket{\Psi(\tau)} \propto c_0 \ket{\varphi_0}+ \sum_{n>0}c_n e^{-(E_n - E_0)\tau}\ket{\varphi_n}\,,
\end{equation}
with $E_i$ the eigenenergies ordered in increasing value. Hence, from~\eqref{eq:appendix_solqite}, it is clear that $\ket{\Psi(\tau)}$ converges to the (non-degenerate) ground state $\ket{\varphi_0}$ of $\hat{H}$ in the limit $\tau \rightarrow\infty$.

Since the imaginary time-evolution is not unitary, obtaining~\eqref{eq:appendix_qiteeq} with a quantum device requires to approximate this action with a unitary operator. A possible way to do this involves approximating the exact ground state with a parametrized trial state,
\begin{equation}~\label{eq:appendix_parametrised_state}
    \ket{\psi(\bm{\theta}(\tau))} =\mathcal{U}(\bm{\theta}(\tau))\ket{\psi_0}\,,
\end{equation}
where $\bm{\theta}\in \mathbb{R}^d$ is a $d-$dimensional parameter vector that depends on the imaginary time. In particular, we will consider parametrised states of the form
\begin{equation}~\label{eq:appendix_parametrised_circuit}
    \mathcal{U}(\bm{\theta}(\tau)) = \prod_\mu e^{-i\theta_\mu (\tau) H_\mu}\,,
\end{equation}
where the different $H_\mu$ are Hamiltonian terms that can be easily generated in the quantum simulator, such as those belonging to the gate set $\mathcal{G}$~\eqref{eq:Gateset} for the case considered here, and $\theta_\mu$ are the elements of $\bm{\theta}$. With such configuration, the Variational Quantum Imaginary Time Evolution (VarQITE) algorithm~\cite{McArdle2019,Yuan2019theoryofvariational,Gacon2024} maps the evolution of the quantum state $\ket{\Psi}$ to the one of the vector of parameters $\bm{\theta}$ using a variational method. 

The application of the McLachan's principle for this particular case, as described in Ref.~\cite{Yuan2019theoryofvariational}, minimizes the distance between the exact imaginary-time wavefunction and the variational one, which is equivalent to the solution of the problem
\begin{equation}~\label{eq:appendix_minimization}
    \delta\left\lVert \left(\frac{\partial}{\partial\tau}+\hat{H}-E_\tau\right)\ket{\psi(\bm{\theta}(\tau)}\right\rVert\,.
\end{equation}
Assuming $\bm{\theta}(\tau)$ to be real,~\eqref{eq:appendix_minimization} becomes
\begin{equation}\label{eq:eqs_motion}
    g\left(\bm{\theta}(\tau)\right) \frac{\mathrm{d}\bm{\theta}}{\mathrm{d}\tau} = \bm{b}\left(\bm{\theta}(\tau)\right)\,,
\end{equation}
which is the projection of~\eqref{eq:appendix_qiteeq} onto the variational manifold. In~\eqref{eq:eqs_motion}, we have introduced the matrix $g = \Re\left[G\right]$ as the real part of the quantum geometric tensor,
\begin{align}\label{eq:G}
    &g_{\mu\nu}\left(\bm{\theta}\right) = \Re\left[\braket{\partial_\mu \psi\left(\bm{\theta}\right) | \partial_\nu \psi\left(\bm{\theta}\right)}\right] \nonumber \\
    & - \Re\left[\braket{\partial_\mu \psi\left(\bm{\theta}\right)| \psi\left(\bm{\theta}\right)}\braket{ \psi\left(\bm{\theta}\right)|\partial_\nu \psi\left(\bm{\theta}\right)}\right]\,,
\end{align}
where we introduce the notation $\partial_\mu := \partial/\partial\theta_\mu$, and the evolution gradient $\bm{b}$,
\begin{equation}\label{eq:b_I}
    b_{\mu} \left(\bm{\theta}\right) =- \Re\left[\braket{\partial_\mu \psi\left(\bm{\theta}\right)|H| \psi\left(\bm{\theta}\right)}\right]\,.
\end{equation}
We note that $b_{\mu} \left(\bm{\theta}\right) = - \partial_\mu E\left(\bm{\theta}\right)/2$, that is, $\bm{b}\left(\bm{\theta}\right)$ is proportional to the gradient of the energy evaluated at $\bm{\theta}$. 

If the parametrized state has enough expressibility (that is, the capability to accurately explore the relevant Hilbert space),~\eqref{eq:appendix_qiteeq} is completely equivalent to~\eqref{eq:eqs_motion}. Although in general this is not the case, ~\eqref{eq:eqs_motion} still provides a good approximation to solve the real-time Schr\"{o}dinger equation if the parametrized state is chosen properly.  The solution of the linear system of ODEs in~\eqref{eq:eqs_motion} involves a feedback loop between the classical and the quantum parts of the algorithm. In practice,~\eqref{eq:eqs_motion} is integrated numerically using an ODE solver. For example, a first order Runge-Kutta method (or Euler method) with a timestep $\Delta \tau$ leads to the following equivalent equation in the limit $\Delta \tau \rightarrow 0$,
\begin{equation}
    g\left(\tau_n\right) \frac{\bm{\theta} \left(\tau_{n+1}\right)-\bm{\theta} \left(\tau_n\right)}{\Delta \tau} = \bm{b}\left(t_n\right)\,,
\end{equation}
where $\tau_n := n\Delta \tau$. Assuming knowledge of the parameters $\bm{\theta}\left(\tau_n\right)$ at the step $n$, the equation above can be solved to yield the following update rule,
\begin{equation}\label{eq:appendix_update_rule}
    \bm{\theta} \left(\tau_{n+1}\right) = \bm{\theta}\left(\tau_n\right) +\Delta \tau\,g^+ \left(\tau_n\right) \bm{b}\left(\tau_n\right)\,,
\end{equation}
where $g^+$ is the pseudoinverse of the metric tensor $g$ (usually regularized adding a small positive constant along the diagonal terms). Therefore, measuring the values of $g$ and $\bm{b}$ at each step $\bm{\theta}(\tau_n)$ is enough to obtain the next set of parameters $\bm{\theta}(\tau_{n+1})$. 

Unfortunately, the measurement of both $g$ and $\bm{b}$ following~\eqref{eq:G} and~\eqref{eq:b_I} involves the measurement of real or imaginary parts of complicated objects as the number of parameters and the imaginary time-evolution increase. In a digital quantum device, this could be solved using controlled unitary operations~\cite{Yuan2019theoryofvariational}; however, in a cold-atom platform, it would be desirable to avoid this overhead. Although there are proposals to do so, such as Ref.~\cite{Yang2024PhaseMeas}, the measurement of $g$ and $\bm{b}$ would add an extra layer of complexity to the algorithm.

To avoid this overhead, and inspired by the non-variational QITE algorithm in Ref.~\cite{motta2020a}, we propose a way to solve~\eqref{eq:appendix_qiteeq} iteratively for each time-step $\Delta\tau$ using the variational approximation. If the state at $\tau_n$ is known, the state at $\tau_{n+1}$ can be obtained exactly as:
\begin{equation}~\label{eq:QITE_step}
    \ket{\Psi\left(\tau_{n+1}\right)} = c(\tau_n) e^{-\Delta\tau \hat{H}}\ket{\Psi\left(\tau_n\right)}\,,
\end{equation}
where $c(\tau_n) = 1/\lVert e^{-\Delta\tau \hat{H}}\ket{\Psi\left(\tau_n\right)} \rVert$ is a normalization constant. Then, taking $\ket{\Psi_0} = \ket{\psi_0} = \ket{\psi(\tau_0)}$ (that is, the exact imaginary-time evolved state at $\tau = 0$ according to~\eqref{eq:appendix_solqite} matches the parametrized quantum state when $\bm{\theta}=0$) and considering parametrized states as the ones in~\eqref{eq:appendix_parametrised_circuit}, then~\eqref{eq:G} and~\eqref{eq:b_I} become
\begin{align}~\label{eq:appendix_gb_decomposition}
    &g_{\mu\nu} \left(\bm{\theta}\left(\tau_0\right)\right)  = \mathrm{Re}\left[\langle \psi_0 |\hat{H}_{\mu}\,\hat{H}_{\nu}|\psi_0\rangle\right]\nonumber\\&-\langle\psi_0|\hat{H}_{\mu}|\psi_0\rangle\,\langle\psi_0|\hat{H}_{\nu}|\psi_0\rangle,\quad \nonumber\\ &b_{\mu} \left(\bm{\theta}\left(\tau_0\right)\right) = \mathrm{Im}\left[\langle \psi_0 |\hat{H}_{\mu}\,\hat{H}|\psi_0\rangle\right]\,,
\end{align}
which can be measured using the methods described in App.~\ref{appendix:measurements}. Hence, from~\eqref{eq:appendix_update_rule}, it is possible to obtain the state at time $\tau_1$, $\ket{\psi(\bm{\tau}_1)}$. Now, based on~\eqref{eq:QITE_step}, we consider this state to be the new initial state, so we map $\ket{\psi(\bm{\tau}_1)}\rightarrow\ket{\psi_0}$ and measure the terms in~\eqref{eq:appendix_gb_decomposition} over this state instead of $\ket{\psi_0}$. The application of the update rule in~\eqref{eq:appendix_update_rule} now gives a new set of parameters that parametrize a quantum circuit generating the state $\ket{\psi(\bm{\tau}_2)}$ from the state $\ket{\psi(\bm{\tau}_1)}$. Applying iteratively this protocol, it is possible to obtain a variational approximation of the imaginary-time evolved quantum state up to $\tau_n$, but always measuring objects of the form described in~\eqref{eq:appendix_gb_decomposition}.

\subsection{Implementing the QLanczos algorithm using the data obtained with the QITE}
\label{app:qlanczos}

Finally, to implementation of the QLanczos algorithm, it is necessary to measure the matrices $\mathbb{H}$ and $\mathbb{S}$, that read
\begin{align}
\mathbb{H}_{\alpha,\beta}&=\bra{\Psi\left(\tau_{\alpha}\right)}\hat{H}\ket{\Psi\left(\tau_{\beta}\right)}\,,\\
\mathbb{S}_{\alpha,\beta}&=\langle\Psi\left(\tau_{\alpha}\right)|\Psi\left(\tau_{\beta}\right)\rangle\,,
\end{align}
where $\ket{\Psi(\tau_\alpha)} = c_\alpha e^{-\alpha\Delta\tau\hat{H}}\ket{\Psi_0}$. To measure these matrices, one can always directly compute each matrix element using the variational approximation of $\ket{\Psi(\tau_\alpha)}$, $\ket{\psi(\bm{\theta}(\tau_\alpha))}$, computing first the absolute value of the term and then its phase. A suitable way to obtain the absolute value would be to use the compute-uncompute method of Refs.~\cite{Gacon2024,Havlicek2017}. This protocol computes the absolute value $\left|\braket{\varphi|\psi}\right|$, where $\ket{\psi}$ is prepared somehow and $\ket{\varphi} = \mathcal{U}\ket{\varphi_0}$, with $\ket{\varphi_0}$ being a product state in the occupation basis (terms of the form $|\braket{\varphi|\hat{H}|\psi}|$ could be computed identically decomposing $\hat{H}$ as a Linear Combination of Unitaries~\cite{Childs2012} and absorbing each unitary into $\ket{\psi}$). To do so, the compute-uncompute method is based on the application of a unitary $\mathcal{V}$ such as $|\braket{\varphi_0|\mathcal{V}|\varphi}| = 1$, so $\bra{\varphi} = e^{i\phi}\bra{\varphi_0}\mathcal{V}$. Then:
\begin{equation}
    \left|\braket{\varphi|\psi}\right|= \left|\braket{\varphi_0 | \mathcal{V} | \psi}\right|\,,
\end{equation}
and now the projection over $\bra{\varphi_0}$ can be done measuring in the occupation basis directly (since it is a product state in that basis). Regarding the operator $\mathcal{V}$, it can be implemented directly as the inverse of $\mathcal{U}$, $\mathcal{U}^\dagger$, by reverting all the signs of the unitaries and its order in the state preparation. However, variational strategies to recompile it with only positive signs can also be applied. On the other hand, once measured the absolute values $\left|\mathbb{H}_{\alpha,\beta}\right|$ and $\left|\mathbb{S}_{\alpha,\beta}\right|$, the estimation of its complex phases could be done following the algorithm described in Ref.~\cite{Yang2024PhaseMeas}.

However, in the main text we also introduce~\eqref{eq:matsHandS} and~\eqref{eq:normalization_approx} to approximately measure these matrices. To derive these approximations we follow the ideas introduced in Ref.~\cite{motta2020a}. First let us note that the normalization constant at step $\gamma$ is
\begin{equation}~\label{eq:appendix_normalization_constant}
    \frac{1}{c_{\gamma}^2} = \braket{\Psi_0|e^{-2\gamma\Delta\tau\hat{H}}|\Psi_0}\,.
\end{equation}
Then, $\mathbb{S}_{\alpha,\beta}$ can be rewritten as
\begin{align}
    \mathbb{S}_{\alpha,\beta} &= \braket{\Psi_0|c_{\alpha}e^{-\alpha\Delta\tau\hat{H}}\,c_{\beta}e^{-\beta\Delta\tau\hat{H}}|\Psi_0} \nonumber\\
    &= c_{\alpha}c_{\beta}\braket{\Psi_0|e^{-(\alpha+\beta)\Delta\tau\hat{H}}|\Psi_0}=\frac{c_{\alpha}c_{\beta}}{c_{\frac{\alpha+\beta}{2}}^2}\,,
\end{align}
and the matrix elements of $\mathbb{H}_{\alpha,\beta}$ read
\begin{align}
    \mathbb{H}_{\alpha,\beta} &= \braket{\Psi_0|c_{\alpha}e^{-\alpha\Delta\tau\hat{H}}\,\hat{H}\,c_{\beta}e^{-\beta\Delta\tau\hat{H}}|\Psi_0} \nonumber\\
    &= c_{\alpha}c_{\beta}\braket{\Psi_0|e^{-\frac{\alpha+\beta}{2}\Delta\tau\hat{H}}\,\hat{H}\,e^{-\frac{\alpha+\beta}{2}\Delta\tau\hat{H}}|\Psi_0}\nonumber\\
    &=\frac{c_{\alpha}c_{\beta}}{c_{\frac{\alpha+\beta}{2}}^2}\braket{\Psi_{\frac{\alpha+\beta}{2}}|\hat{H}|\Psi_{\frac{\alpha+\beta}{2}}}=\mathbb{S}_{\alpha,\beta}\,\braket{\Psi_{\frac{\alpha+\beta}{2}}|\hat{H}|\Psi_{\frac{\alpha+\beta}{2}}}\,.
\end{align}
Hence, all the matrix elements can be computed using the normalization constants and measurements of the energy. Finally, the normalization constants can be computed recursively noticing that $c_0 = 1$ and, from~\eqref{eq:appendix_normalization_constant}:
\begin{align}
    \frac{1}{c_{\gamma+1}^{2}} &= \braket{\Psi_0|e^{-2(\gamma+1)\Delta\tau\hat{H}}|\Psi_0}=\frac{\braket{\Psi_\gamma|e^{-2\Delta\tau\hat{H}}|\Psi_\gamma}}{c_{\gamma}^2}\nonumber\\
    &=\frac{1-2\Delta\tau\braket{\Psi_\gamma|\hat{H}|\Psi_\gamma}}{c_{\gamma}^2}+\mathcal{O}(\Delta\tau^2)\,,
\end{align}
where in the last line we have expanded the exponential up to first order in $\Delta\tau$.
\bibliographystyle{apsrev4-2}
\bibliography{references_2,referencesalex}

\end{document}